\documentclass[final,3p,times,authoryear]{elsarticle}

\usepackage{mymacro}
\usepackage{amsmath}
\usepackage{amssymb}

\usepackage{lineno}
\usepackage{rotating}
\usepackage{color}
\usepackage{caption}
\usepackage{subcaption}

\journal{Icarus}

\begin{document}

\begin{frontmatter}



\title{Modeling the disequilibrium species for Jupiter and Saturn: Implications for Juno and Saturn entry probe}


\author{Dong Wang}
\ead{dw459@cornell.edu}
\address{Department of Astronomy, 610 Space Sciences Building, Cornell University,
    Ithaca, NY 14853}

\author{Jonathan I. Lunine}
\ead{jlunine@astro.cornell.edu}
\address{Department of Astronomy, 610 Space Sciences Building, Cornell University,
    Ithaca, NY 14853}

\author{Olivier Mousis}
\address{Aix Marseille Universit\'e, CNRS, LAM (Laboratoire d'Astrophysique de Marseille) UMR 7326, 13388, Marseille, France}

\begin{abstract}
Disequilibrium species have been used previously to probe the deep water abundances and the eddy diffusion coefficient for giant planets. In this paper, we present a diffusion-kinetics code that predicts the abundances of disequilibrium species in the tropospheres of Jupiter and Saturn with updated thermodynamic and kinetic data. The dependence on the deep water abundance and the eddy diffusion coefficient is investigated. We quantified the disagreements in CO kinetics that comes from using different reaction networks and identified C$_2$H$_6$ as a useful tracer for the eddy diffusion coefficient. We first apply a H/P/O reaction network to Jupiter and Saturn's atmospheres and suggest a new PH$_3$ destruction pathway. New chemical pathways for SiH$_4$ and GeH$_4$ destruction are also suggested, and another AsH$_3$ destruction pathway is investigated thanks to new thermodynamic and kinetic data. These new models should enhance the interpretation of the measurement of disequilibrium species by JIRAM on board Juno and allow disentangling between methods for constraining the Saturn's deep water abundance with the Saturn entry probes envisaged by NASA or ESA.   
     
\end{abstract}

\begin{keyword}
Abundances, atmospheres -- Jupiter, atmosphere -- Saturn, atmosphere


\end{keyword}

\end{frontmatter}


\section{Introduction}
Disequilibrium species in the atmosphere of Jupiter and Saturn can be used to constrain the deep water abundance and the deep eddy diffusion coefficient in the atmospheres of Jupiter and Saturn \citep[e.g.,][]{PB77, FL94}. Various disequilibrium species such as CO, PH$_3$, GeH$_4$, and AsH$_3$ have been detected on Jupiter and Saturn with abundances orders of magnitude higher than their respective chemical equilibrium abundances at the pressure level where they are observable \citep[e.g.,][]{Beer75, Noll86, Ridgway76, Larson80, Fink78, Noll88b, Bezard89, Noll89}. These species at a few bars are transported upward by vertical mixing from the deep atmosphere where they are more abundant, therefore, they contain the information of the atmosphere down to a few hundred bars. In this paper, we model the vertical profiles of disequilibrium species with updated thermodynamic and kinetic data. The dependence on the water abundance and the eddy diffusion coefficient is investigated. 

Our study is timely for the following reasons. The JIRAM instrument on board the Juno spacecraft will be able to measure the disequilibrium species CO, PH$_3$, GeH$_4$, and AsH$_3$ down to a few bars when it will arrive at Jupiter in 2016 \citep{Grassi10}. The microwave radiometer onboard Juno will also be able to measure the deep water abundance \citep{Janssen05}. With the abundances of disequilibrium species and water, constraints should be made on the deep eddy diffusion coefficient. A Saturn probe proposal has been submitted to the ESA 2015 call for medium class mission \citep{Mousis15} and a similar concept is under study for a submission to the NASA 2016 New Frontier call\citep{Atkinson12}. Current entry probes are designed to go down to 10--20 bar and can make in-situ measurements of the atmosphere composition via mass spectrometry \citep{Wong04}. However, it is unlikely that such probes will be able to descend below the water cloud deck and measure the deep water abundance. A study with the updated kinetic data is then necessary for evaluating whether deep water abundance can be effectively constrained by disequilibrium species. 

We use the diffusion-kinetic model developed in \citet{Wang15}. A C/N/O/H reaction network is employed to predict the abundances of various carbon bearing species. The reaction networks for P/H/O and Si/H/O species are applied for the first time to study planetary atmospheres in this paper. New chemical pathways for PH$_3$ and GeH$_4$ destructions are then proposed. New compilations of thermochemical data, especially for P, Ge, and As, are used in our model. 

The paper is organized as follows. In section 2, we introduce the current status of the measurement of disequilibrium species. In section 3, we describe our models for the chemistry and transport of disequilibrium species. In section 4, we present our results. In section 5, we discuss the implications for Juno and a Saturn entry probe. The conclusions are summarized in section 6.  

\section{Measurements of disequilibrium species: current status}
The tropospheric abundances of CO, PH$_3$, SiH$_4$, GeH$_4$ and AsH$_3$ are primarily measured in the 5$\mu$m window for Jupiter and Saturn. Apart from a 1 ppb tropospheric component \citep[e.g.,][]{Larson78, Bjoraker86}, CO also has a stratospheric component \citep[e.g.,][]{Bezard02}. The tropospheric CO is supplied by vertical convective mixing from deep levels where CO prevails \citep{PB77}, while the stratospheric CO can be supplied by micrometeoroids \citep{Prather78}, infalling materials from icy satellites \citep{Strobel79}, or shock chemistry from infalling kilometer to subkilometer-sized comets \citep{Lellouch95, Bezard02}. At Jupiter and Saturn, comets are more probable than other sources \citep{Bezard02,Cavalie10}. The tropospheric CO contains information on the deep atmosphere, and thus can be used to probe the deep water abundance and the deep eddy diffusion coefficient. The retrieval of Saturn's tropospheric CO has not been successful due to its very low mixing ratios \citep{Cavalie09}. Tropospheric PH$_3$ was measured in the 5$\mu$m window for Jupiter with a mixing ratio of $(6\sim9) \times 10^{-7}$ \citep[e.g.,][]{Kunde82, Bjoraker86, Encrenaz96, Irwin98}, and for Saturn with a mixing ratio of $(3\sim5) \times 10^{-6}$ \citep[e.g.,][]{NL91, deGraauw97, Fletcher11}. The vertical profile of PH$_3$ was retrieved from the spectra by Cassini CIRS and VIMS \citep{Fletcher09a, Fletcher11}. The PH$_3$ abundance starts to be depleted in the upper troposphere where the pressure is about 1 bar \citep{Fletcher12}, due to decreased eddy mixing, UV photolysis and chemical re-equilibration \citep{Irwin98,Irwin04}. Tropospheric GeH$_4$ was identified and measured at the 5$\mu$m window \citep{Fink78, Noll88b} at a mixing ratio of a few times 10$^{-10}$. The mixing ratio of GeH$_4$ in the stratosphere is expected to be lower than that in the troposphere because of UV photolysis. The tropospheric AsH$_3$ was measured on both Jupiter and Saturn \citep{Bezard89, Noll89, NL91}. The mixing ratio of AsH$_3$ on Jupiter is about $2\times10^{-10}$ \citep{Noll90}, while for the Saturn, the mixing ratio is about $3\times10^{-9}$ \citep{Bezard89, NL91}. In table \ref{tab: compositions}, we summarize the measurements of tropospheric CO, PH$_3$, SiH$_4$, GeH$_4$, and AsH$_3$ abundances for both Jupiter and Saturn. 

\section{Model}

\subsection{Introduction to the model}

We developed a code to solve the 1-D transport-kinetic equation: 
\begin{equation}\label{eqn: TK}
\frac{\partial{Y_i}}{\partial{t}} = \frac{1}{\rho} \frac{\partial}{\partial{z}}(\rho K_{\rm eddy} \frac{\partial{Y_i}}{\partial{z}}) + P_i - L_i,
\end{equation} 
where $Y_i$ is the mass fraction of species i, $\rho$ is the density of the atmosphere, $z$ is the vertical coordinate relative to a reference point in the atmosphere (we choose the 1 bar level in the code), $K_{\rm eddy}$ is the vertical eddy diffusion coefficient, $P_i$ is the 
chemical production rate of species i, and $L_i$ is the chemical loss rate of species i. Both $P_i$ and $L_i$ have a unit of g cm$^{-3}$ s$^{-1}$. The time evolution of $Y_i$ is controlled by two physical processes: one is the chemical production and destruction of species i, and the other is its corresponding vertical transport. In the convective envelope of Jupiter and Saturn, the transport of mass is mainly by turbulent convection. Here in equation (\ref{eqn: TK}), the convective transport of species is approximated by diffusion transport with an coefficient $K_{\rm eddy}$, which is a good approximation justified by the success of mixing length theory in explaining stellar convection \citep{Stone76}. The mass fractions $Y_i$ are initialized using their local chemical equilibrium values along the adiabat. The chemical net production rate ($P_i - L_i$) is integrated using \textit{Cantera}, a software toolkit developed for problems involving chemical kinetics and thermodynamics \citep{Cantera15}. At each time step, we call \textit{Cantera} to do the integration and include the result in the resolution of the continuity equation. \textit{Cantera} has been used and tested for many applications including combustion, detonation, fuel cells, batteries, etc. The integration is terminated when the mass fractions $Y_i$ reach steady state. The code requires three kinds of input. One is the temperature pressure profile ($T-P$ profile), the second is a list of thermodynamic properties in the format of NASA polynomials \citep{McBride93} for each species, and a list of reactions between these species, the third is the elemental composition and the vertical eddy diffusion coefficient $K_{\rm eddy}$. 

\begin{figure}
\begin{center}
\resizebox{\hsize}{!}{\includegraphics[angle=0]{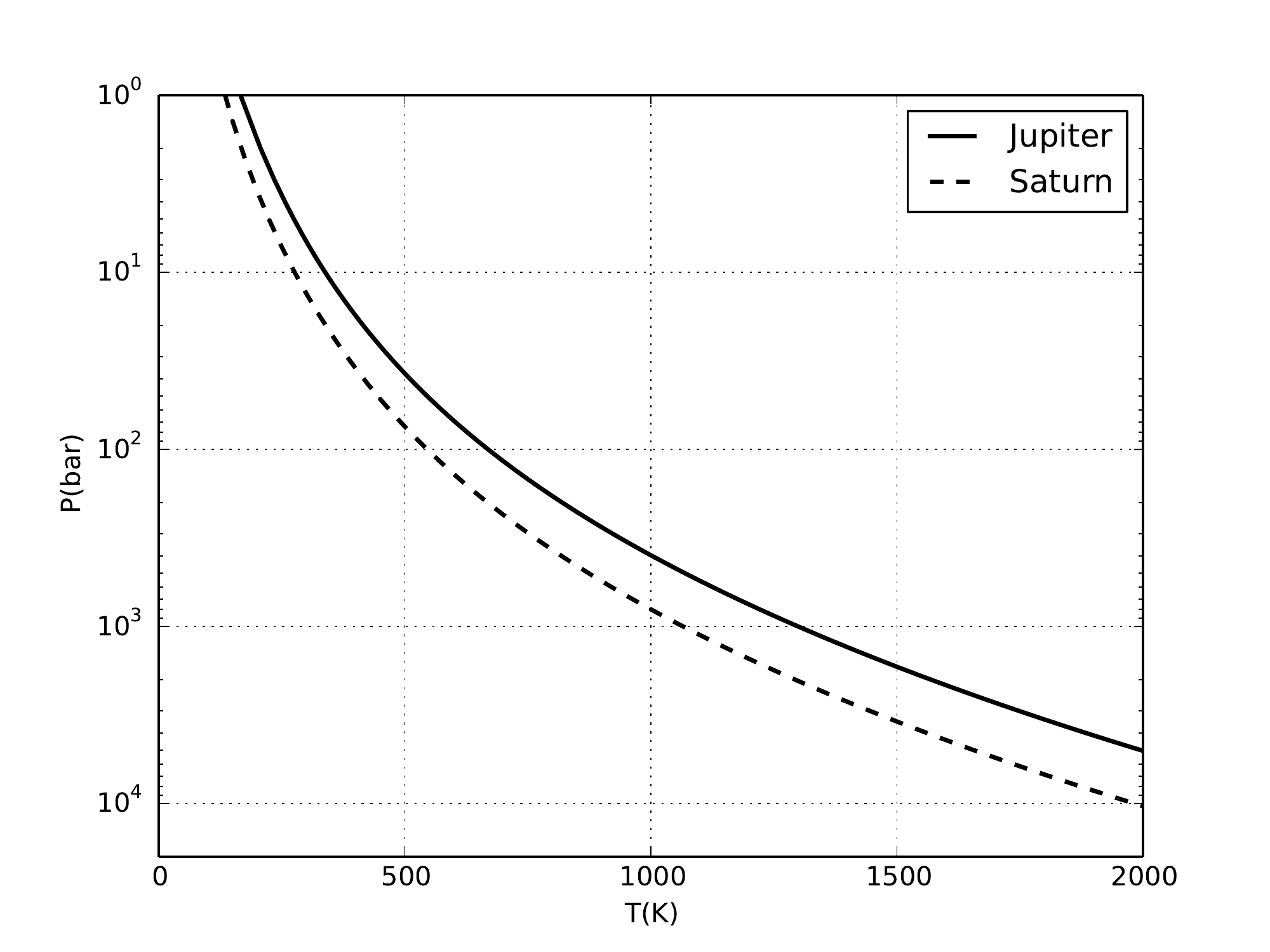}}
\caption{Adiabatic profiles of Jupiter and Saturn computed following the method described in \citet{FP85}.}
\label{fig: adiabat}
\end{center}
\end{figure}

The $T-P$ profiles for Jupiter and Saturn are calculated following the method described in \citet{FP85}. We choose a reference point where temperature and pressure are measured and extrapolated into the deep atmosphere assuming a dry adiabat. For Jupiter, we use $T$ = 427.71 K at 22 bars as our reference point \citep{Seiff98}. The pressure and temperature conditions lower than 22 bars are from the Galileo entry probe measurements \citep{Seiff98}. For Saturn, we use $T$ = 134.8 K at $P$ = 1 bar as our reference point \citep{Lindal85}. The heat capacity of the atmosphere used in the calculation is computed by linearly combining the heat capacities of H$_2$ and He, and the heat capacities of H$_2$ and He are from NIST-JANAF thermochemical table \citep{Chase98}. The helium mixing ratio we use is 0.157 for Jupiter \citep{Niemann98} and 0.135 for Saturn \citep{CG00}. The T-P profiles for Jupiter and Saturn are calculated and shown in Fig. \ref{fig: adiabat}. 
The thermodynamic and reaction data are gathered from various sources which are detailed in the following:   
\begin{itemize}
\item{\it C/N/O/H reaction network}
 
Our C/N/O/H reaction network used in this paper is developed based on the network from \citet{Venot12} downloaded from the KIDA database \citep[][\underline{http://kida.obs.u-bordeaux1.fr}]{Wakelam12}. The network consists of 105 neutral species and 963 reactions. Among the reactions, 957 of them are reversible reactions and 6 of them are irreversible reactions. A complete list of the species can be found in \citet{Venot12}. The network has been validated against various combustion experiments in the temperature range between 300 K and 2000 K and in the pressure range between 0.01 bar and several hundred bars. 

Alternative network applied to the hydrogen-rich atmospheres is that of \citet{Moses11} and \citet{VM11}. \citet{Moses14} compared their model with the \citet{Venot12} model and found that the major difference in CO/CH$_4$ chemistry comes from the rate coefficient of the reaction H + CH$_3$OH $\leftrightarrow$ CH$_3$ + H$_2$O. The \citet{Venot12} model used the rate coefficient obtained by \citet{Hidaka89} from laboratory experiments. However, \citet{Moses14} argued that this reaction likely is prohibited by a very large energy barrier and is much slower than was estimated by \citet{Hidaka89} based on quantum chemical calculations in \citet{Moses11}. It remains to be seen whether changing the rate coefficient following the suggestions by \citet{Moses11} would reproduce the experimental results in \citet{Hidaka89}. Therefore, because the discrepancy remains unresolved, we have considered two reaction networks in our model, which are: 
\begin{itemize}
\item{\it network A}: it is based on the \citet{Venot12} network with some modifications. Among the species in the list, we remove HNC because it does not participate in any reactions in the reaction network. We include CH$_3$HN$_2$, CH$_3$HN, CH$_2$NH2, and CH$_2$NH into the network since these species are expected to be important in a hydrogen rich environment \citep{Moses10}. The final network A consists of 108 species and 1000 reactions. Added reactions and their rate coefficients are from \citet{DB00}. The thermodynamic properties are mainly compiled from \citet{BR05}, \citet{McBride93}, \citet{DB00} and \citet{Venot12}. An online updated version of the \citet{BR05} database can be found at \underline{http://garfield.chem.elte.hu/Burcat/burcat.html}.
The whole reaction list along with thermodynamic data and rate coefficient data are available in the KIDA database (\underline{http://kida.obs.u-bordeaux1.fr/networks.html}).   

\item{\it network B}: it is the same as the network A except the rate coefficient for the reaction H + CH$_3$OH $\leftrightarrow$ CH$_3$ + H$_2$O is revised to be much slower following the recommended rate in \citet{Moses11}. The slower rate leads to nearly two orders of magnitude increase in the CO/CH$_4$ conversion timescale near the quench level. 

\end{itemize}
     
\item{\it H/P/O reaction network}

The H/P/O reaction network is based on \citet{Twarowski95}. The reaction network consists of 24 species and 175 reactions. The phosphorus containing species included are: PH$_3$, PH$_2$, PH, HOPO, HPO, PO, 
PO$_2$, PO$_3$, HOPO$_2$, P$_2$O$_3$, P, P$_2$, P$_4$, P$_2$O, P$_2$O$_2$, HPOH, H$_2$POH. The network has been used to explain the faster recombination rate of H and OH in the presence of phosphine combustion products \citep{Twarowski96}. To make sure all important species under Jupiter/Saturn's atmospheric conditions are included, we performed an equilibrium calculation using the NASA Chemical Equilibrium Application (CEA) \citep{GM94,MG96} for temperature and pressure conditions along the adiabats of Jupiter and Saturn. We find H$_3$PO$_4$ is important but missing from this reaction network, so we added it and associated reactions into the reaction network. The thermodynamic data are primarily from \citet{BR05}, \citet{McBride93}. The thermodynamic data for P$_2$O and P$_2$O$_2$ are from \citet{Twarowski93}. The whole reaction network is available in the KIDA database.
  
\item{\it Si/O/H reaction network}

The Si/O/H reaction network is from \citet{Miller04}. The network consists of 69 species and 198 reactions. The silicon bearing species included in the network are: 
                    Si, Si$_2$, Si$_3$, SiH, SiH$_2$, 
                     cis-OSiH$_2$O, Si$_2$H$_2$, SiH$_4$, SiH$_3$, H$_3$SiSiH$_3$,
                     H$_3$SiSiH,    H$_2$SiSiH$_2$,   Si$_2$H$_5$,      Si$_2$H$_3$,      Si$_2$O$_2$,
                     Si$_3$O$_3$,      Si$_4$O$_4$,      Si$_5$O$_5$,      Si$_6$O$_6$,      Si$_7$O$_7$,
                     Si$_8$O$_8$,      Si$_9$O$_9$,      Si$_{10}$O$_{10}$,    (SiH$_2$O)$_2$,   SiO$_2$,
                     H$_3$SiOSiH$_3$,  H$_3$SiOOH,    H$_3$SiOO,     SiOOH,      H$_2$SiOH,
                     H$_3$SiO,      HOSiO$_2$,     SiO,        Si(OH)$_2$,    SiOH,
                     H$_2$Si(OH)O,  H$_3$SiOH,     HSiOH,      HSiO,       H$_2$SiO,
                     HSiO(OH),   HSiO$_2$,      HOSiO,      HSiOOH,     SiO$_2$(c),
                     SiO$_2$(l),    SiO$_2$(g),    H$_2$SiOOH,    (HSiOOH)$_2$,  Si$_2$O$_4$,
                     Si$_3$O$_6$,      Si$_4$O$_8$,      Si$_5$O$_{10}$,     Si$_6$O$_{12}$,     Si$_7$O$_{14}$,
                     Si$_8$O$_{16}$,     Si$_9$O$_{18}$,     Si$_{10}$O$_{20}$. 
This network has been used to model the combustion of silane (SiH$_4$).  The whole reaction network is available in the KIDA database. 
\end{itemize}

For arsenic (As) and germanium (Ge) containing species, there is no reaction network available in the literature. We use the quenching timescale approach to compute the abundances of GeH$_4$ and AsH$_3$ along Jovian and Saturnian adiabats. This approach does not need a whole reaction network but a single rate determining step. This method was widely used in the literature as an alternative to the diffusion-kinetic modeling \citep[e.g.,][]{PB77, FL94}, and can correctly predict the quench level as long as the appropriate rate determining step is used and the recipe in \citet{Smith98} is followed to compute the mixing timescale. A detailed description of this approach can be found in \citet{Wang15}. 

For equilibrium computations, we use the NASA Chemical Equilibrium Application (CEA) as our reference \citep{GM94, MG96}. This code is in wide use in the aerodynamic and thermodynamic community. The code includes over 2000 species in the database.

The elemental abundances of Jupiter and Saturn used in this paper are summarized in table \ref{tab: elements}. We use the solar composition table from \citet{Asplund09} as our reference. The elemental abundances are inferred from the observed abundances of their hydrogen compounds. For phosphorus, we assume the total phosphorus abundance is equal to the observed PH$_3$ abundance. For silicon and germanium, we assume they have enrichments similar to carbon, since their observed hydrogen compounds (SiH$_4$ and GeH$_4$) do not represent their total elemental abundances. For arsenic, we assume its total abundance is represented by the observed AsH$_3$ abundance.   

\subsection{Validation of the Diffusion-Kinetic Code}

\begin{figure}
\begin{center}
\resizebox{\hsize}{!}{\includegraphics[angle=0]{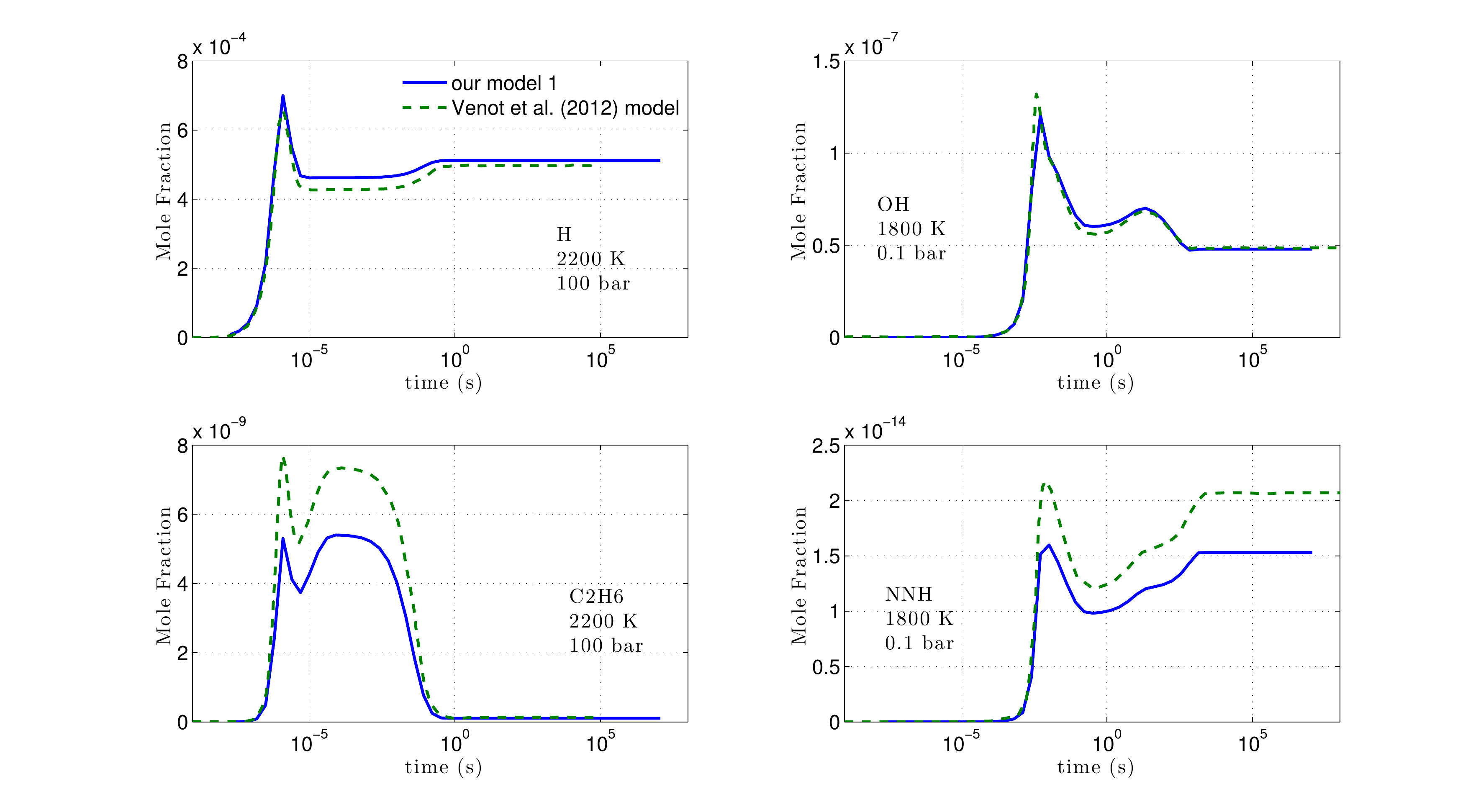}}
\caption{Time evolution of the chemical species H, OH, C$_2$H$_6$ and NNH under constant temperature and pressure. Solid lines are computed using our code, and dashed lines are from Fig. 1 in \citet{Venot12}. The initial condition is a mixture of gas H$_2$, He, O$_2$, CH$_4$, and N$_2$ with solar elemental abundances. }
\label{fig: 0_D}
\end{center}
\end{figure}

We first test our code without diffusion using the C/N/O/H reaction network computing the time evolution of the abundances of different species at constant temperature and pressure. The integration is done using \textit{Cantera}. A comparison between our results and those from the nominal model of \citet{Venot12} is shown in Fig. \ref{fig: 0_D}. The evolution is very similar for H and OH, but not exactly the same for C$_2$H$_6$ and NNH. The difference is due to the utilized thermodynamic properties. Indeed, we have changed the thermodynamic data to those provided in \citet{Venot12} and exactly obtained the same evolution of the abundance profiles for all four species. This prompts us to compare the thermodynamic data we used with those in \citet{Venot12thesis}. Our thermodynamic data are gathered from widely used compilations, for example, \citet{McBride93} and \citet{BR05}. The species that are not available from literature are estimated using the software THERM \citep{Ritter91}. \citet{Venot12} also gathered thermodynamic data in a similar manner. However, we do not know the source of thermodynamic data for each species in \citet{Venot12}, therefore, comparisons for individual species is not possible. This is not expected to be a major source of uncertainty since the uncertainties in the kinetic data are much larger than those in the thermodynamic data.  
 
\begin{figure}
\begin{center}
\resizebox{\hsize}{!}{\includegraphics[angle=0]{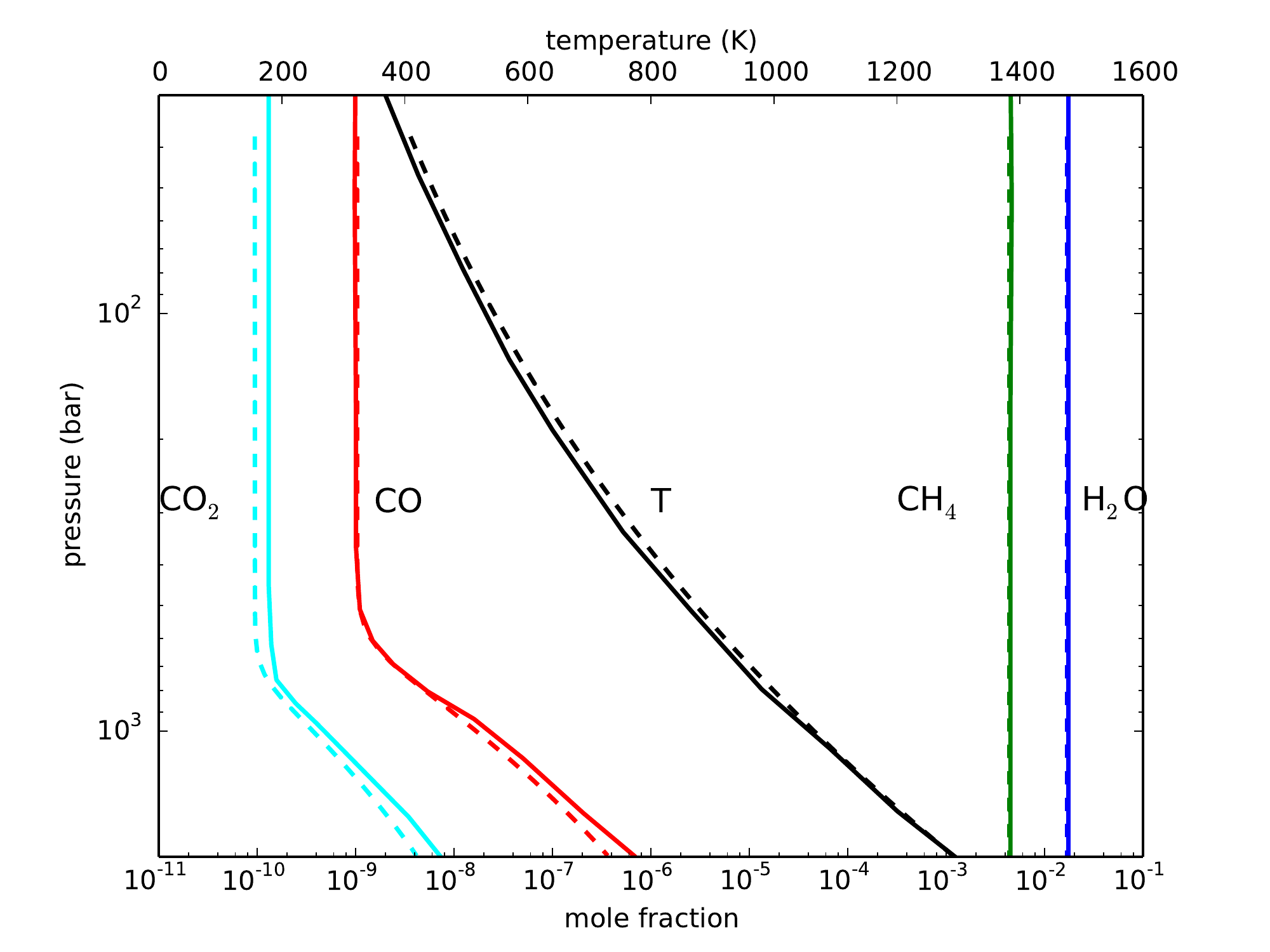}}
\caption{Comparison between our model and \citet{Venot12} model on mole fraction profiles along Saturn's adiabat. 
Solid lines are from Fig. 1 in \citet{Mousis14}, which is computed using the \citet{Venot12}'s model. Dashed lines are computed using our model. The elemental abundances we used here are O/H = 21 times solar, C/H = 9 times solar, and $K_{\rm eddy}$ = 1$\times$10$^9$ cm$^2$ s$^{-1}$, the same as those used in Fig. 1 of \citet{Mousis14}. 
}
\label{fig: 1_D}
\end{center}
\end{figure}

The test with both kinetics and diffusion is done by simulating the Saturn's atmosphere thermochemistry using the C/N/O/H reaction network, and comparing against the results in Fig. 1 of \citet{Mousis14}. The reference result in \citet{Mousis14} for Saturn is computed using the same code as in \citet{Venot12}. The comparison, represented in Fig. \ref{fig: 1_D},  shows that the differences in the mixing ratios are within 10$\%$. There might be three sources of error that contribute to the differences of the mixing ratios shown in Fig. \ref{fig: 1_D}: (1) the temperature-pressure profile; (2) the thermodynamic data; (3) the elemental abundances, which are all inputs to our code. These difference are small and the comparison shows our code can correctly solve the diffusion-kinetics Equation (\ref{eqn: TK}).   

\section{Results}
In this section, we predict the abundances of various disequilibrium species in Jupiter's and Saturn's atmospheres at a few bars pressure. Above this level altitude, the abundances are affected by photochemistry and the enrichment from external sources, which are not included in our calculations here. The dependence on the water abundance and the deep eddy diffusion coefficient are investigated. 

We make a few definitions regarding the abundance of species Z. Its concentration is denoted as [Z] with units of molecules$\cdot$cm$^{-3}$. Mole fraction of Z is denoted as $X_{\rm Z}$ = [Z]/$n$, where $n$ is total number density of the atmosphere (molecule$\cdot$cm$^{-3}$). Mixing ratio is denoted as $q_{\rm Z}$ = [Z]/[H$_2$]. For an element M, we define $E_{M}$ as the enrichment of M relative to solar, which is the ratio of [M]/[H] in the planet to that in the Sun, where [M] and [H] are the total number density of M and H atoms respectively, in whatever form. The elemental composition of the solar atmosphere is taken from \citet{Asplund09}. The elemental abundances of Jupiter and Saturn used in our simulations can be found in Table \ref{tab: elements}. 

\subsection{Simulation results using the C/N/O/H reaction network}

\begin{figure}
\begin{center}
\resizebox{\hsize}{!}{\includegraphics[angle=0]{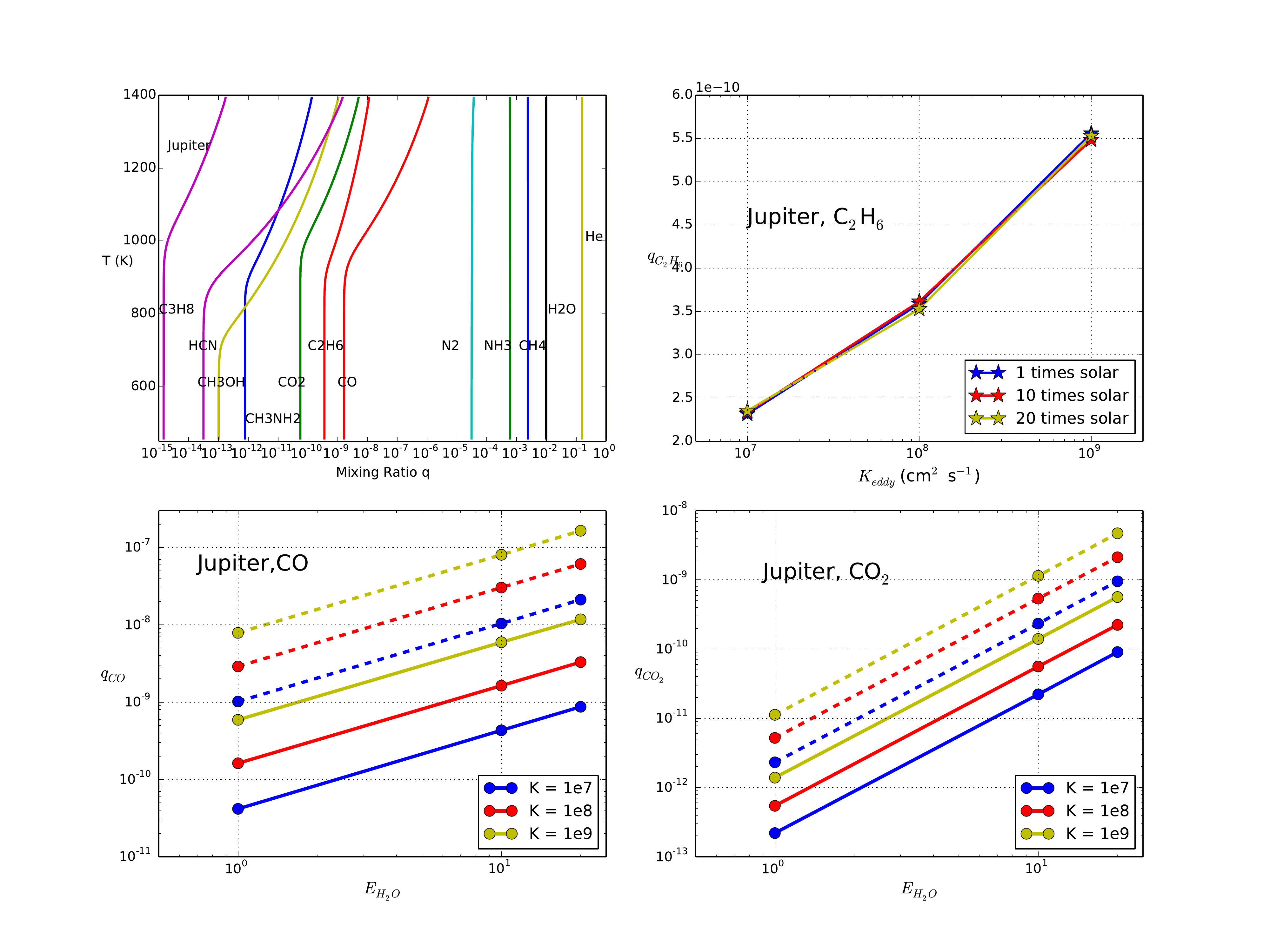}}
\caption{Simulation results for Jupiter using the C/N/O/H reaction network. Upper left plot is the mixing ratio of species along Jupiter's adiabatic profile for parameters: $E_{\rm H_2O}$ = 10 and $K_{\rm eddy}$=1$\times$10$^8$ cm$^2$s$^{-1}$ using the network A. Only species with mixing ratios $q>1\times10^{-15}$ at 450 K are plotted. 
Upper right plot is the predicted mixing ratio of C$_2$H$_6$ as a function of $K_{\rm eddy}$ for a range of assumed water abundances. Solid lines are calculated using network A, and dashed lines are calculated using network B. 
Lower plots are the predicted mixing ratio of CO and CO$_2$ as a function of the water enrichment $E_{\rm H_2O}$ for different vertical eddy diffusion coefficient $K_{\rm eddy}$. 
Solid lines are calculated using network A, and dashed lines are calculated using network B.}
\label{fig: carbon_jupiter}
\end{center}
\end{figure}

The C/N/O/H reaction network used is described in section 2. We considered the two aforementioned reaction networks, namely, network A and network B. The vertical profile of the mixing ratios of species is computed and the ($E_{\rm H_2O}$, $K_{\rm eddy}$) parameter space is investigated. As an illustration, the mixing ratios along Jupiter's adiabat for parameters $E_{\rm H_2O}$ = 10 and $K_{\rm eddy}$=1$\times$10$^8$ cm$^2$s$^{-1}$ using the network A are presented in Fig.  \ref{fig: carbon_jupiter}. Our calculations show that N$_2$ is the major nitrogen bearing species after NH$_3$ with a mixing ratio of about 1 ppm. C$_2$H$_6$, CO and CO$_2$ are the major carbon bearing species after CH$_4$. CO and CO$_2$ are the major oxygen bearing species after H$_2$O.  The mixing ratio of CO and C$_2$H$_6$ are at 1 ppb level, while the mixing ratio of CO$_2$ is about 0.1 ppb. Mixing ratios of other species such as CH$_3$NH$_2$ and HCN are below 1$\times$10$^{-12}$.

We focus on three species, CO, C$_2$H$_6$ and CO$_2$, and investigate their dependence on parameters $E_{\rm H_2O}$ and $K_{\rm eddy}$. For each  combination of ($E_{\rm H_2O}$, $K_{\rm eddy}$), we run the simulation to steady state. The mixing ratios of C$_2$H$_6$, CO, and CO$_2$ have been extracted from each simulation and the results are summarized in Fig. \ref{fig: carbon_jupiter}. As is shown by the figure, the abundance of C$_2$H$_6$ is sensitive to the vertical eddy diffusion coefficient while it is not affected by the deep water abundance. Therefore, C$_2$H$_6$ is a good tracer for the eddy diffusion coefficient alone. In contrast, the abundance of CO is both sensitive to the eddy diffusion coefficient and the deep water abundance. The constraints on the deep water abundance placed by the 
CO abundance is limited by the information on the eddy diffusion coefficient. If both CO and C$_2$H$_6$ can be measured, we can use the mixing ratio of C$_2$H$_6$ to put a tight constraint on the $K_{\rm eddy}$, then we can determine how much water is needed to match the observed CO abundance with the predicted CO abundance. 
CO$_2$ is another tracer for the deep water abundance and the eddy diffusion coefficient. Although it is very sensitive to the water abundance ($\propto E_{\rm H_2O}^2$), its mixing ratio is an order of magnitude less than CO, making it more difficult to be measured. In Fig. \ref{fig: carbon_jupiter}, we also show the results computed using the network B. The predictions on CO and CO$_2$ abundances using network B are very different from those using network A. Therefore, significant uncertainties still remain for the CO kinetics that require laboratory measurements for their definite resolutions.  

The tropospheric CO abundance on Jupiter was measured at the Northern Equatorial Belt ($\sim 9^{\circ}$) by \citet{Bezard02} with a mole fraction of $(1.0\pm0.2)\times10^{-9}$ . Combined with the predicted eddy diffusion coefficient of $\sim 1\times10^8$ cm$^2$s$^{-1}$ \citep{Wang15} near the equatorial regions, the deep water abundance is constrained to be $\sim 7$ times solar with network A and $\sim 0.6$ times solar with network B. 

The simulation results for Saturn are presented in Fig. \ref{fig: carbon_saturn}. Similarly to the results for Jupiter, we find C$_2$H$_6$ is a good tracer for the eddy diffusion coefficient. Both CO and CO$_2$ are good tracers for the deep water abundance. Dual constraints from CO and C$_2$H$_6$ can break the degeneracy between high (low) eddy diffusion coefficient and high (low) deep water abundance.  

The tropospheric CO abundance on Saturn was not measured so far and an upper limit is put by \citet{Cavalie09} with a mixing ratio smaller than 1 ppb. Combined with the predicted eddy diffusion coefficient of $\sim 1\times10^8$ cm$^2$s$^{-1}$ \citep{Wang15} near the equator, the deep water abundance is constrained to be $\lesssim 60$ times solar with network A and $\lesssim 10$ times solar with network B.   

\subsection{Simulation results using the H/P/O reaction network}

We use the NASA CEA chemical equilibrium code to compute the equilibrium abundances of phosphorus containing species along Saturn's adiabat, and the results are shown in Fig. \ref{fig: P_eq}. The elemental abundances used in the calculations are summarized in Table \ref{tab: elements}. Our calculations show that H$_3$PO$_4$ is the major phosphorus bearing species below 700 K instead of P$_4$O$_6$ in the literature \citep[e.g.,][]{FL94, VF05}. We find the difference is due to the different thermodynamic data used for P$_4$O$_6$. \citet{FL94} and others use the standard enthalpy of formation of P$_4$O$_6$ ($\Delta H_f^{0} [\rm P_4O_6]$) from JANAF table \citep{Chase85}, which is based on the experiments by \citet{KD52}. As a comparison, the CEA code uses the $\Delta H_f^{0} [\rm P_4O_6]$ from Gurvich's table \citep{Gurvich90}, which is based on the experiments by \citet{HM63}. \citet{FL94} favored the data from the JANAF table because they see ``no compelling reasons, such as a problem with experimental methods or data reduction, to reject the work of \citet{KD52} in favor of the work of \citet{HM63}". They point out that the discrepancy needs to be resolved by a new experimental determination of the $\Delta H_f^{0} [\rm P_4O_6]$. However, we favor the data from Gurvich's table for the following reasons. \citet{HM63} actually pointed out that in the experiment of \citet{KD52} the sample examined is a mixture of P$_4$O$_6$ and P$_4$O$_{10}$ rather than pure P$_4$O$_6$. The identification of P$_4$O$_6$ is not necessarily definitive and it may lead to the more negative values of the enthalpy of formation for P$_4$O$_6$ (due to pollution by P$_4$O$_{10}$). Later studies by \citet{Muenow70} and \citet{Smoes73} also supported the measurement by \citet{HM63}. Quantum chemical calculations by \citet{Morgon12} get values of the enthalpy of formation of P$_4$O$_6$ that are closer to the values of \citet{HM63}. We didn't find another direct or indirect measurement that supported the values by \citet{KD52}. The Burcat database \citep{BR05} adopted the values by \citet{HM63} and pointed out that the values used in JANAF is erroneous. For these reasons, we choose to take the data from Gurvich's table. Our computation shows that PH$_3$ is converted to H$_3$PO$_4$ at  about 700 K in Saturn's atmosphere. From our computations for Jupiter, PH$_3$ is largely converted to H$_3$PO$_4$ at about 650 K in Jupiter's atmosphere. 

The main chemical pathway for PH$_3$/H$_3$PO$_4$ conversion is identified from the H/P/O reaction network by comparing the rates of all the reactions in the network. This method is very robust since no information is needed other than the kinetic data. The details of our analysis are presented in the Appendix B. We find the main chemical pathway consists of the following reactions: 

\begin{subequations}\label{eqn: P_pathway}
\begin{align}
& \textrm{PH}_{3}  \leftrightarrow \textrm{PH}_2 + \textrm{H}    \\
& \textrm{PH}_{2} +  \textrm{H}_{2}\textrm{O} \leftrightarrow \textrm{H}_{2}\textrm{POH} + \textrm{H} \\
& \textrm{H}_{2}\textrm{POH} + \textrm{PH}_{2} \leftrightarrow \textrm{H}\textrm{POH} + \textrm{PH}_{3} \\
& \textrm{H}\textrm{POH}  \leftrightarrow \textrm{H}\textrm{PO} + \textrm{H} \\  
& \textrm{H}\textrm{PO}  \leftrightarrow \textrm{PO} + \textrm{H}   \\
& \textrm{PO} +  \textrm{H}_{2}\textrm{O}  \leftrightarrow \textrm{HOPO} + \textrm{H} \\
& \textrm{HOPO} + \textrm{H}  \leftrightarrow \textrm{PO}_{2} + \textrm{H}_{2} \\ 
& \textrm{PO}_{2} + \textrm{H}_{2}\textrm{O}  \leftrightarrow \textrm{HOPO}_{2} + \textrm{H} \\
& \textrm{HOPO}_{2} + \textrm{H}_{2}\textrm{O}  \leftrightarrow \textrm{H}_{3}\textrm{PO}_{4} \\
\cline{1-2}
& \textrm{PH}_{3} + 4\, \textrm{H}_{2}\textrm{O}  \leftrightarrow \textrm{H}_{3}\textrm{PO}_{4} + 4\textrm{H}_{2} \tag{\ref{eqn: P_pathway}, net} 
\end{align}
\end{subequations}

The rate determining step for the pathway is reaction (\ref{eqn: P_pathway}h).  
The rate coefficient for this reaction is unknown, however, the backward reaction coefficient is estimated in \citet{Twarowski95}. For the reverse reaction of \ref{eqn: P_pathway}h, the rate coefficient $k_{\ref{eqn: P_pathway}h,r} = 5.24\times10^{-11}e^{-6014/T}$ cm$^3$molecule$^{-1}$s$^{-1}$. Using the detailed balance, the forward rate coefficient $k_{\ref{eqn: P_pathway}h} = k_{\ref{eqn: P_pathway}h,r} K_{\ref{eqn: P_pathway}h, eq}$, where $K_{\ref{eqn: P_pathway}h, eq}$ is the equilibrium constant of reaction (\ref{eqn: P_pathway}h). Therefore, the forward rate coefficient is estimated to be $k_{\ref{eqn: P_pathway}h} = 2.35\times10^{-12} e^{-1.067\times10^4/T}$ cm$^3$molecule$^{-1}$s$^{-1}$. 

In Figures \ref{fig: PH3_J} and \ref{fig: PH3_S}, we have plotted the predicted mixing ratios of  PH$_3$ as a function of the $K_{\rm eddy}$ computed using the rate determining step. Our calculations show that the mixing ratio of PH$_3$ is not sensitive to the values of $K_{\rm eddy}$ unless $K_{\rm eddy}$ is less than 1$\times$10$^5$ cm$^2$s$^{-1}$. We also compared the predicted PH$_3$ for different values of $E_{\rm H_2O}$. The mixing ratio of PH$_3$ is also not
sensitive to the value of $E_{\rm H_2O}$ unless $E_{\rm H_2O}$ is greater than 30. PH$_3$'s insensitivity to parameters is due to its quench level deep in the regime where it is the dominant species. For example, when $K_{\rm eddy}$ = 1$\times$10$^7$ cm$^2$s$^{-1}$ and $E_{\rm H_2O} = 10$, the quench level  is at $T \approx 900 K$. From our equilibrium calculations shown in Fig. \ref{fig: P_eq}, PH$_3$ is stable and dominant until below 700 K. Almost all the phosphorus is sequestered in PH$_3$ on Jupiter and Saturn. 

\subsection{Simulation results for SiH$_4$}

Figure \ref{fig: Si_eq} shows the predicted equilibrium abundances of some Si-bearing species along Saturn's adiabat computed using the NASA CEA code. The elemental abundances used in the calculations are summarized in Table \ref{tab: elements}. MgSiO$_3$ (l,s) and Mg$_2$SiO$_4$(s) condensate are the primary carriers of Si below 2000 K. 
The mixing ratio of  SiH$_4$ is 1$\times$10$^{-9}$ at about 1500 K, and decreases to 1$\times$10$^{-18}$ at about 1000 K. The equilibrium results are in agreement with the 
results from \citet{FL94}. Observationally, the upper limit for $q_{\rm SiH_4}$ on Jupiter is 2.5$\times$10$^{-9}$ \citep{Treffers78},
and the upper limit for $q_{\rm SiH_4}$ on Saturn is 2.0$\times$10$^{-10}$ \citep{NL91}. This indicates that almost all of the silicon on Jupiter
and Saturn is removed by the rock formation and subsequent condensation. The total oxygen in Jupiter should be at least the oxygen locked in the rocks \citep[e.g.,][]{VF05}. Since MgSiO$_3$(s) is the major condensates for silicon, each silicon atom is combined with three oxygen atoms. Assuming the total silicon enrichment relative to solar is similar to the enrichment for carbon, then Si/H is about 1.43$\times$10$^{-4}$ on Jupiter and 3.24$\times$10$^{-4}$
on Saturn. The oxygen locked with silicon therefore has a mixing ratio of O/H $>$ 4.4$\times$10$^{-4}$ for Jupiter and O/H $>$ 1.0$\times$10$^{-3}$ for Jupiter and Saturn, corresponding to about 0.9 and 2.0 times solar, respectively. Adding this part of oxygen and the oxygen sequestered in water would give the total oxygen in Jupiter and Saturn. 

Thanks to the detailed kinetic data of \citet{Miller04}, we can extract the main chemical pathway for SiH$_4$ destruction from the reaction network. The main chemical pathway is identified by comparing the rates of all the reactions in the network. The analysis is detailed in the Appendix C. The main chemical pathway of SiH$_4$ destruction is given by the following reactions:  

\begin{subequations}\label{eqn: Si_pathway}
\begin{align}
& \textrm{SiH}_{4}  + \textrm{M} \leftrightarrow \textrm{SiH}_{2} + \textrm{H}_{2} + \textrm{M},    \\
& \textrm{SiH}_{2} +  \textrm{H}_{2}\textrm{O} \leftrightarrow \textrm{HOSiH} + \textrm{H}_2, \\
& \textrm{HSiOH} \leftrightarrow \textrm{SiO} + \textrm{H}_{2}, \\
& \textrm{SiO} + \textrm{H}_{2}\textrm{O} \leftrightarrow \textrm{Si(OH)}_2,       \\  
& \textrm{Si(OH)}_2 + \textrm{H}  \leftrightarrow \textrm{HOSiO} + \textrm{H}_2,   \\
& \textrm{Si(OH)}_2 \leftrightarrow \textrm{HSiO(OH)}   \\
& \textrm{HSiO(OH)} + \textrm{H}  \leftrightarrow \textrm{HOSiO} + \textrm{H}_2,   \\
& \textrm{HOSiO}   \leftrightarrow \textrm{SiO}_2(cr) + \textrm{H}, \\
& \textrm{SiO}_2(cr) + \textrm{Mg(OH)}_2  \leftrightarrow \textrm{MgSiO}_{3}(s) + \textrm{H}_{2}\textrm{O}, \\
\cline{1-2}
& \textrm{SiH}_{4} + \textrm{H}_{2}\textrm{O} + \textrm{Mg(OH)}_2 \leftrightarrow \textrm{MgSiO}_{3}(s) + 4\textrm{H}_2 \tag{\ref{eqn: Si_pathway}, net}. 
\end{align}
\end{subequations}  

The steps (\ref{eqn: Si_pathway}a) to (\ref{eqn: Si_pathway}h) are directly identified from the network, but the last step (\ref{eqn: Si_pathway}i) is added because SiO$_2$(cr) is not a stable product in the atmospheres of Jupiter and Saturn according our equilibrium calculations and those of \citet{FL94}. A reaction incorporating Si into to MgSiO$_3$(s) completes the chemical pathway.   
Compared with the chemical path proposed in \citet{FL94}, the formation of SiH$_2$ and MgSiO$_3$(s) is the same, but the path from SiH$_2$ to SiO$_2$(cr) is different. The difference is due to the methodologies used to identify main chemical pathways. \citet{FL94} constructed their chemical pathway by identifying important intermediates observed in the experiments and then devising a pathway to connect the reactants, intermediates, and the products. There might be other important intermediates that are missing from the pathway, and the way of connection is not unique. This method is used when we do not have a reaction network or do not have enough kinetic data to perform detailed analysis. However, since we now have a reaction network, we can simulate the chemical evolution using the network and identify the fastest route from SiH$_4$ to SiO$_2$(cr). Our analysis is detailed in the Appendix C.           
Among the chemical pathway, we find reactions \ref{eqn: Si_pathway}b, \ref{eqn: Si_pathway}e, and \ref{eqn: Si_pathway}g all serve as the bottlenecks. There is no unique rate limiting step for SiH$_4$ destruction. 

This analysis helps us simplify the original H/Si/O reaction network from \citet{Miller04}. The original network is designed for the simulation of SiH$_4$ burning in the molecular oxygen. Therefore, the composition is silicon-rich and oxygen-rich. However, the atmosphere of Jupiter and Saturn is hydrogen rich. Some species in the reaction network are expected to be unimportant in Jupiter and Saturn's atmosphere, and thus can be removed from the network to increase the efficiency of
time integration. In the simplified reaction network, the species included are H$_2$, He, H$_2$O, OH, H, SiH$_4$, SiH$_3$, SiH$_2$, SiO, Si(OH)$_2$, HSiOH, HSiO(OH), HOSiO, and SiO$_2$(c). These species are the major intermediates for SiH$_4$ destruction, as is shown in Fig. \ref{fig: path_Si} in the Appendix C. This simplified reaction network is used to     
predict the abundance of SiH$_4$ along Saturn's adiabat for different values of $K_{\rm eddy}$ and $E_{\rm O}$. The results are presented in Fig. \ref{fig: SiH4}. For $E_{\rm O} = 2$ and $K_{\rm eddy}$ = 1$\times$10$^9$ cm$^2$ s$^{-1}$, the predicted mixing ratio of SiH$_4$ at a few bars level is about $1\times10^{-17}$, while for $E_{\rm O} = 30$ and $K_{\rm eddy}$ = 1$\times$10$^7$ cm$^2$ s$^{-1}$, the mixing ratio of SiH$_4$ falls to 1$\times$10$^{-26}$. For the two extreme cases considered here, SiH$_4$ abundance is too small to be detected. Therefore, we conclude SiH$_4$ is not expected in Saturn's troposphere. The same analysis is applied to Jupiter. Similar to the result for Saturn, SiH$_4$ is also not expected in Jupiter's troposphere. \citet{FL94}'s calculation for SiH$_4$ kinetics also concludes the low mixing ratio of SiH$_4$. We confirm this result using a new Si/H/O reaction network. 

\subsection{Simulation results for GeH$_4$}
In Fig. \ref{fig: Ge_eq}, we show the equilibrium abundances of germanium containing species along Saturn's adiabat computed using the NASA CEA code. The elemental abundances used in the calculations are summarized in Table \ref{tab: elements}.
The figures show that GeH$_4$ is not the dominating germanium containing species, and it is converted to more abundant GeS at lower temperature. Condensation happens near 700 K, and the major condensates are Ge(cr), GeS(cr) and GeO$_2$(cr). There are limited data for the GeH$_4$ kinetics, therefore, we propose a chemical pathway for the conversion between GeH$_4$ and GeS by analogy with SiH$_4$ and SiO, since Si and Ge are in the same group and next to each other in the periodic table, and O and S are in the same group and also next to each other in the periodic table. The main chemical pathway for the conversion from SiH$_4$ to SiO is given by reactions \ref{eqn: Si_pathway}a, 
\ref{eqn: Si_pathway}b, and \ref{eqn: Si_pathway}c. The rate determining reaction is \ref{eqn: Si_pathway}b. By analogy, we propose the main chemical pathway below for the conversion from GeH$_4$ to GeS.  

\begin{subequations}\label{eqn: Ge_pathway}
\begin{align}
& \textrm{GeH}_{4}  + \textrm{M} \leftrightarrow \textrm{GeH}_{2} + \textrm{H}_{2} + \textrm{M},    \\
& \textrm{GeH}_{2} +  \textrm{H}_{2}\textrm{S} \leftrightarrow \textrm{HSGeH} + \textrm{H}_2, \\
& \textrm{HSGeH} \leftrightarrow \textrm{GeS} + \textrm{H}_{2},
\end{align}
\end{subequations}

and the rate determining reaction is \ref{eqn: Ge_pathway}b. This chemical pathway is very similar to the one proposed in \citet{FL94}. The only difference is the step from GeH$_2$ to HSGeH. In \citet{FL94}, two steps are used: the first is GeH$_2$ + H$_2$S $\leftrightarrow$ H$_2$Ge=S + H$_2$, and the second is H$_2$Ge=S  $\leftrightarrow$ HGe=SH. In the H/Si/O reaction network, the rate coefficient for SiH$_2$ + H$_2$O $\leftrightarrow$ H$_2$Si=O + H$_2$ is $3.84\times10^{10} T^{-0.6} e^{-4905/T}$ cm$^3$mol$^{-1}$s$^{-1}$, and the rate coefficient for SiH$_2$ + H$_2$O $\leftrightarrow$ HSi=OH + H$_2$ is 2.15$\times10^{10} T^{0.7} e^{-4956/T}$ cm$^3$mol$^{-1}$s$^{-1}$ \citep{ZT95}. At 800 K, the first reaction is much
slower than the second reaction. By analogy, we therefore favor the single step from GeH$_2$ to HGe=SH proposed here instead of the double step proposed in \citet{FL94}.  

The rate coefficient for the reaction \ref{eqn: Ge_pathway}b is not in the literature. Again by analogy with reaction SiH$_2$ + H$_2$O, we use $k_{\ref{eqn: Ge_pathway}b} \approx 2.15\times10^{10}T^{0.7}e^{-4956/T}$ cm$^3$mol$^{-1}$s$^{-1}$ \citep{ZT95}. The predicted mole fraction of GeH$_4$ as a function of the vertical eddy diffusion coefficient $K_{\rm eddy}$ is shown in Fig. \ref{fig: GeH4_J} and \ref{fig: GeH4_S} for Jupiter and Saturn, respectively. For Jupiter, our predicted mole fraction is consistent with the observed mixing ratio of $7^{+4}_{-2} \times 10^{-10}$ by \citet{Bjoraker86} with the values of $K_{\rm eddy}$ predicted by Wang et al. (2015). For Saturn, our predicted mole fraction is also consistent with the observed value of 
(4$\pm$2)$\times$10$^{-10}$ by \citet{Noll88b}. This consistency indicates the rate coefficient of GeH$_2$ + H$_2$S is close to our estimate here. Within a factor 5 uncertainty on both sides of the rate coefficient (total factor of 25), we find the predicted mixing ratios of GeH$_4$ is in agreement with the observed values. Adjusting the rate coefficient even higher or lower will yield a disagreement between the prediction and the observation. Therefore, the rate coefficient of reaction \ref{eqn: Ge_pathway}b should be within a factor of 5 of our estimation. When we have an accurate measurement  of the rate coefficient, we can use the GeH$_4$ abundance to constrain the value of $K_{\rm eddy}$. The sensitivity of GeH$_4$ abundance on the $K_{\rm eddy}$ implies the effectiveness of using GeH$_4$ abundances to constrain the $K_{\rm eddy}$, but the kinetics needs to be better constrained by laboratory measurements first.    

\subsection{Simulation results for AsH$_3$} 

Equilibrium calculations show that AsH$_3$ is the dominant arsenic bearing species until it is converted to As$_4$ or As$_2$S$_2$ at about 400 K \citep{FL94}. 
Due to the lack of kinetic data, it is quite uncertain where the quench level is. \citet{FL94} proposed three possible chemical pathways of the conversion from AsH$_3$ to As$_4$ or As$_2$S$_2$. The first pathway starts from the combination of AsH and AsH$_3$ forming As$_2$H$_2$, then As$_2$H$_2$ is decomposed into As$_2$ which combines to form the As$_4$ condensates. A similar mechanism starts from two AsH$_2$ forming the As$_2$H$_2$, and the rest is the same. The third one start from the combination of AsH with HS forming AsS, and two AsS combines to form the As$_2$S$_2$ condensates. The first and the third chemical pathway were investigated by \citet{FL94}. The second one was not investigated because the authors do not have the thermodynamic data for AsH$_2$. However, the second pathway is more likely than the fist pathway since AsH$_2$ is expected to be more abundant than AsH. Now we study the second pathway, which is described by the following reactions: 

\begin{subequations}\label{eqn: As_pathway}
\begin{align}
& \textrm{AsH}_{3}  + \textrm{H} \leftrightarrow \textrm{AsH}_{2} + \textrm{H}_{2},    \\
& \textrm{AsH}_{2} +  \textrm{AsH}_{2} \leftrightarrow \textrm{As}_2\textrm{H}_2 + \textrm{H}_2, \\
& \textrm{As}_2\textrm{H}_2 \leftrightarrow \textrm{As}_2 + \textrm{H}_{2}, \\
& \textrm{As}_2 + \textrm{As}_2 \leftrightarrow \textrm{As}_4, \\
& \textrm{As}_4 \leftrightarrow \textrm{As}_4(s), 
\end{align}
\end{subequations} 

where the rate determining step is taken as the reaction \ref{eqn: As_pathway}b. The thermodynamic data for AsH$_2$ is from \citet{TP86}. The rate coefficient of the reaction \ref{eqn: As_pathway}b is taken as $4.63\times10^{-11} T^{0.04} e^{-16.8/T}$ cm$^3$ molecule$^{-1}$ s$^{-1}$, by analogy with the rate coefficient of SiH$_2$ + SiH$_2$ \citep{Dollet07}. The analogy between As and Si is probably not good, but we are forced by the lack of kinetic data for elements in the same column. We consider an overall two order of magnitude uncertainty for the rate coefficient. The kinetic results are shown in Fig. \ref{fig: AsH3_J} and \ref{fig: AsH3_S}.  From our calculations, the predicted AsH$_3$ abundance is 
nearly equal to the total arsenic abundance for a large range of the $K_{\rm eddy}$. Now with the three chemical pathways studied, the conclusion is that AsH$_3$ abundance is not sensitive to the value of $K_{\rm eddy}$, and AsH$_3$/H $\approx$ As/H in both Jupiter and Saturn.  

The AsH$_3$ is observed on Jupiter with a mixing ratio of $(2.2\pm1.1)\times10^{-10}$ \citep{Noll90}, corresponding to 0.3$\sim$0.8 times solar abundance. The AsH$_3$ is observed on Saturn with a mixing ratio of $(3\pm1)\times10^{-9}$ \citep{Bezard89, Noll89, NL91}, corresponding to 5$\sim$10 times solar abundance. The higher enrichment of As on Saturn than on Jupiter is consistent with other elements, such as C and P. However, the subsolar As/H ratio on Jupiter is puzzling because other rock forming elements such as phosphorus and germanium are all enriched relative to solar \citep[e.g.,][]{FL94}. There could be three possibilities. One is the observations have underestimated the AsH$_3$ abundance in Jupiter, and the AsH$_3$ abundance could be higher than solar. The second possibility is the observations are correct, but the kinetics are not. This may be hard to understand since the kinetics work well for Saturn. The third possibility is both observations and the kinetics are correct, and the subsolar abundance of As is realistic, but then a mechanism is needed to deplete arsenic on Jupiter. The JIRAM instrument on board Juno spacecraft has the capability of measuring the AsH$_3$ abundance at a few bars in Jupiter's atmosphere \citep{Grassi10}. However, to resolve this issue, experimental measurements on AsH$_3$ are necessary to better determine the arsenic kinetics.
  
\section{Discussion}
In section 4, we have presented our modeling results on the disequilibrium species using updated thermodynamic and kinetic data for Jupiter and Saturn. We have explored the dependence on the two free parameters in our model: the eddy diffusion coefficient and the deep water abundance. We find that CO and CO$_2$ are sensitive to both the eddy diffusion coefficient and the deep water abundance, while C$_2$H$_6$ and GeH$_4$ are only sensitive to the eddy diffusion coefficient. In this section, we discuss how Juno and a Saturn probe can improve our understanding on the $K_{\rm eddy}$ and $E_{\rm H_2O}$ through the measurement of disequilibrium species. We also discuss the uncertainties on the kinetic networks and suggest reactions that should be further studied. 
 
\subsection{Further constraints by Juno}
Juno is about to arrive at Jupiter in July 2016, and the microwave radiometer is expected to map the water abundance down to $\sim$100 bars. The JIRAM instrument will measure the abundances of H$_2$O, NH$_3$, PH$_3$, CO, GeH$_4$, AsH$_3$ at a few bars level \citep{Grassi10}. Using the measured deep water abundance and the tropospheric CO abundance measured by \citet{Bezard02}, the eddy diffusion coefficient can be constrained as is shown by Fig. \ref{fig: carbon_jupiter} in this paper and Fig. 5 in \citet{Wang15}. Another constraint on the $K_{\rm eddy}$ can be placed using the measured GeH$_4$ abundance, as is shown by Fig. \ref{fig: GeH4_J}. However, a quantified reaction network is needed, or at least a determination of the rate limiting reaction is required to reduce the uncertainties in GeH$_4$ kinetics. 

Both \citet{FG78} and \citet{Wang15} predicted latitudinal variations on the deep eddy diffusion coefficient due to the rotational effects on the convection. The latitudinal variation of the eddy diffusion coefficient should result in latitudinal variations of the abundances of CO, C$_2$H$_6$, CO$_2$, and GeH$_4$ at a few bars level. Multi-latitude measurements of CO and GeH$_4$ will test the prediction, if performed by JIRAM. We predict no latitudinal variation on the PH$_3$ and AsH$_3$ abundances since their abundances are not sensitive to the eddy diffusion coefficient. The horizontal (latitudinal) profile of CO on Jupiter is presented in the Fig. 9 of \citet{Wang15}. The mole fraction of CO is predicted to be about $1\times10^{-9}$ near the equator and decreases to about $4\times10^{-10}$ near the pole. Fig. 10 of Wang et al. (2015) shows latitudinal variations of PH3 on Jupiter because the old chemical model is used in that paper. With the new chemical model for phosphorus in this paper, we predict no latitudinal variations for PH$_3$ at a few bars level. The predicted horizontal profile of GeH$_4$ is shown in Fig. \ref{fig: GeH4_J_horizontal}. The horizontal profile of eddy diffusion coefficient used in the calculations is from \citet{Wang15}. The mole fraction of GeH$_4$ is predicted to be about 7$\times$10$^{-10}$ near the equator and slowly decreases to about 3$\times$10$^{-10}$ near the pole. 

\subsection{Application to a Saturn entry probe} 
\begin{figure}
\begin{center}
\resizebox{\hsize}{!}{\includegraphics[angle=0]{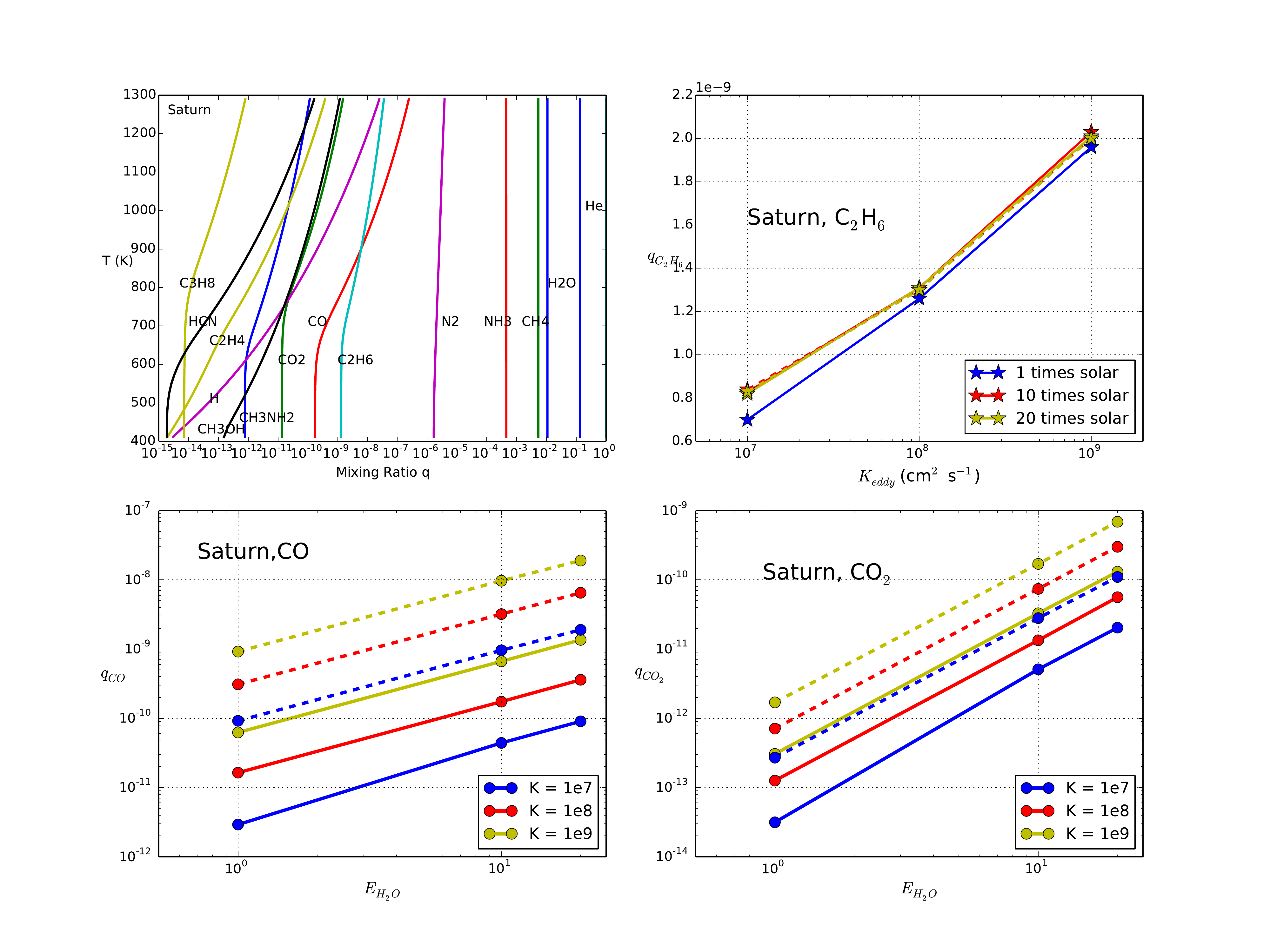}}
\caption{Simulation results for Saturn using the C/N/O/H reaction network. Upper left plot is the mixing ratio of species along Saturn's adiabatic profile for parameters: $E_{\rm H_2O}$ = 10 and $K_{\rm eddy}$=1$\times$10$^8$ cm$^2$s$^{-1}$ using the network A. Only species with mixing ratios $q>1\times10^{-15}$ at 400 K are plotted. 
Upper right plot is the predicted mixing ratio of C$_2$H$_6$ as a function of $K_{\rm eddy}$ for a range of assumed water abundances. Solid lines are calculated using network A, and dashed lines are calculated using network B. 
Lower plots are the predicted mixing ratio of CO and CO$_2$ as a function of the water enrichment $E_{\rm H_2O}$ for different vertical eddy diffusion coefficient $K_{\rm eddy}$. 
Solid lines are calculated using network A, and dashed lines are calculated using network B.}
\label{fig: carbon_saturn}
\end{center}
\end{figure}

\begin{figure}
\begin{center}
\resizebox{\hsize}{!}{\includegraphics[angle=0]{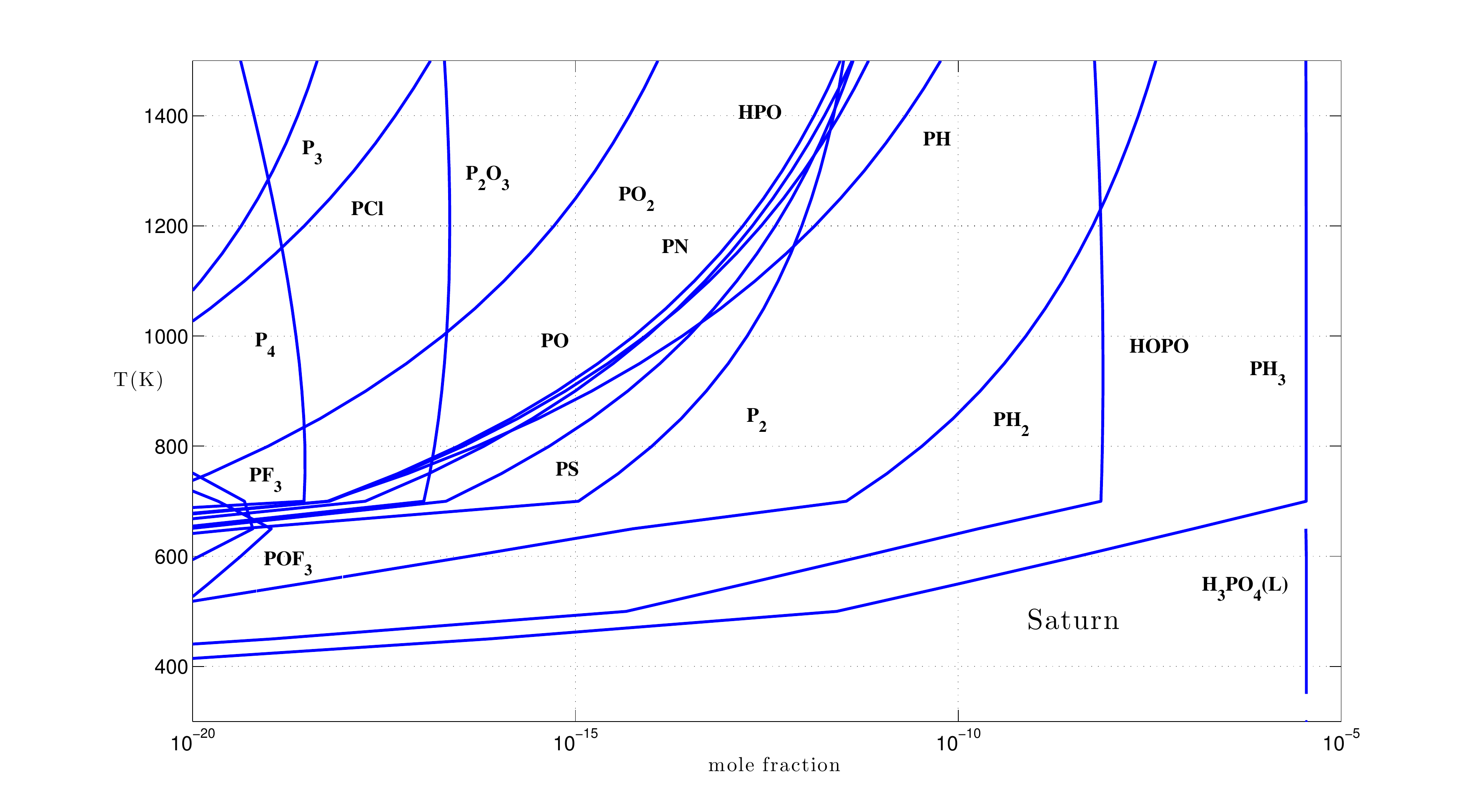}}
\caption{Equilibrium mole fractions of phosphorus containing species along Saturn's adiabat computed using the NASA Chemical Equilibrium Application (CEA) software.
The elemental input we used are P/H = 7.8 times solar, O/H = 10 times solar, and other elemental abundances are solar except He, C, N, S, Si, Ge, and As, which are listed in table \ref{tab: elements}. } 
\label{fig: P_eq}
\end{center}
\end{figure}

\begin{figure}
\begin{center}
\resizebox{\hsize}{!}{\includegraphics[angle=0]{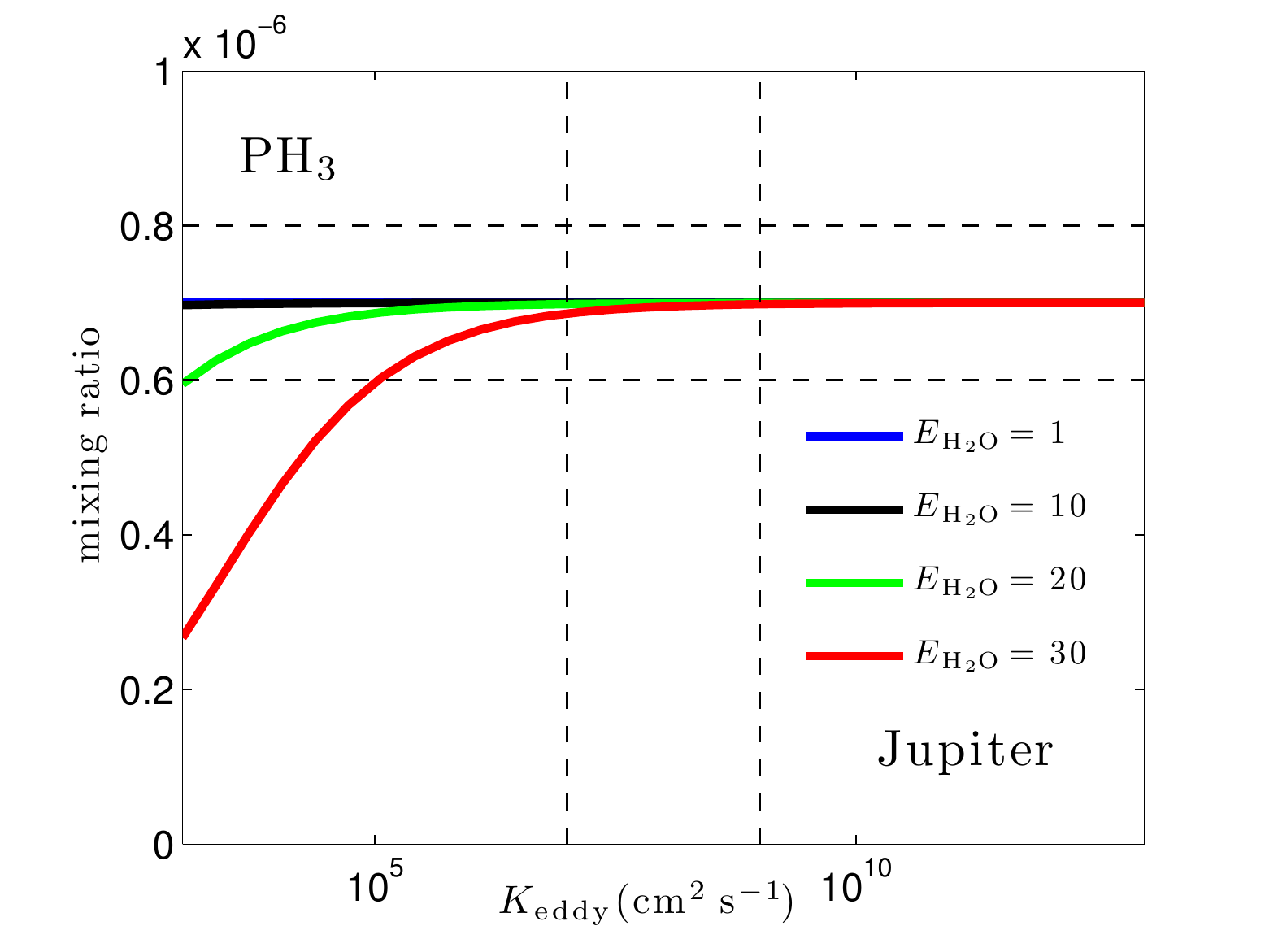}}
\caption{The predicted mixing ratio of PH$_3$ as a function of the vertical eddy diffusion coefficient $K_{\rm eddy}$ in Jupiter's atmosphere at a few bars level. The horizontal dashed lines show the range of observed PH$_3$ mixing ratio $q_{\rm PH_3}=7.0\pm1.0\times$10$^{-7}$ \citep[e.g.,][]{Irwin98}. 
The vertical dashed lines show the range of plausible eddy diffusion coefficient.} 
\label{fig: PH3_J}
\end{center}
\end{figure}

\begin{figure}
\begin{center}
\resizebox{\hsize}{!}{\includegraphics[angle=0]{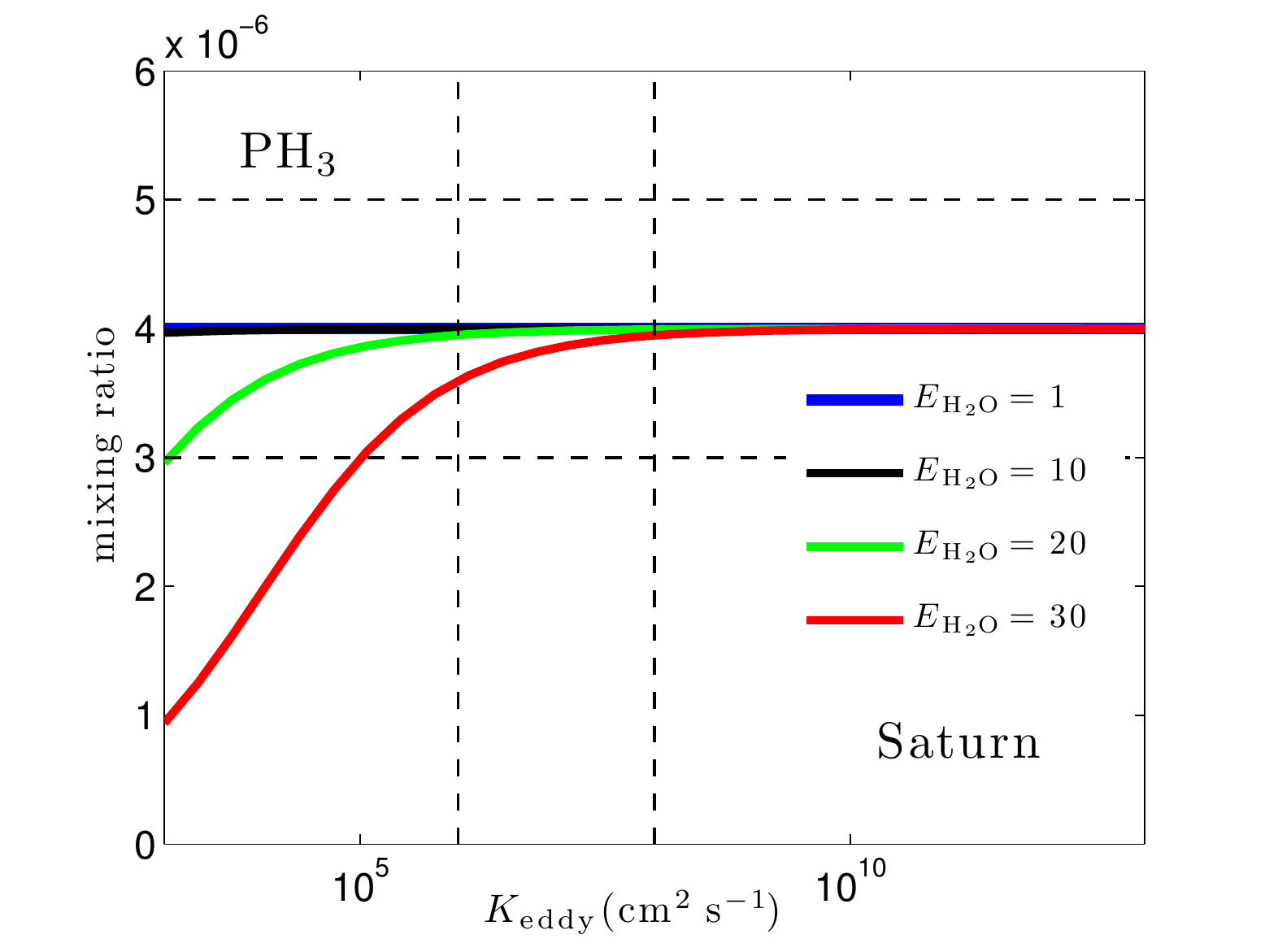}}
\caption{The predicted mixing ratio of PH$_3$ as a function of the vertical eddy diffusion coefficient $K_{\rm eddy}$ in Saturn's atmosphere at a few bars level.The horizontal dashed lines show the range of observed PH$_3$ mixing ratio $q_{\rm PH_3}=4.0\pm1.0\times$10$^{-6}$ \citep[e.g.,][]{Fletcher11}. 
The vertical dashed lines show the range of plausible eddy diffusion coefficient.} 
\label{fig: PH3_S}
\end{center}
\end{figure}

\begin{figure}
\begin{center}
\resizebox{\hsize}{!}{\includegraphics[angle=0]{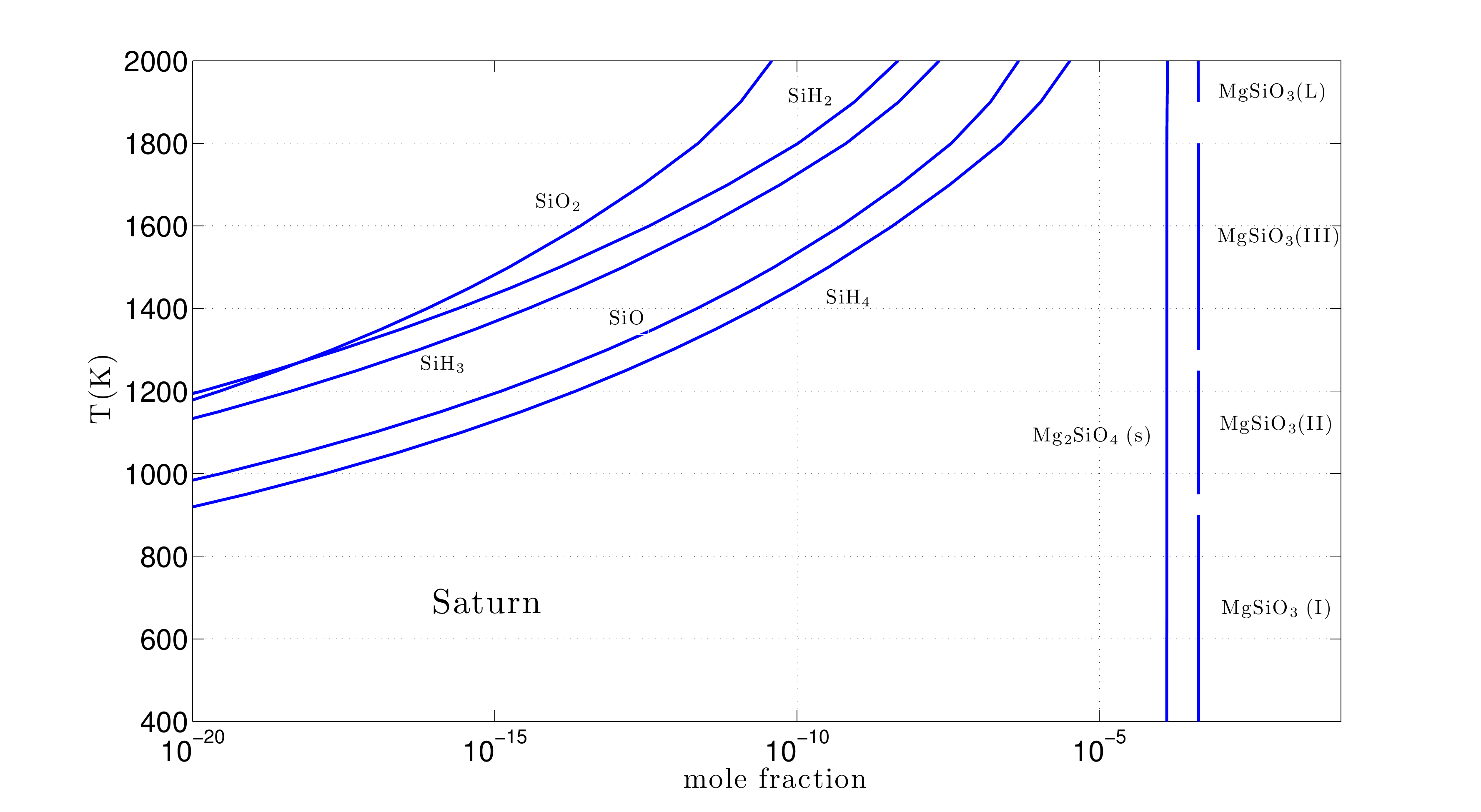}}
\caption{Equilibrium mole fractions of some Si-bearing species along Saturn's adiabat computed using the NASA CEA code. The elemental input we used are Si/H = 10 time solar, O/H = 10 times solar, and other elemental abundances are solar except He, C, N, S, P, Ge, and As, which are listed in table \ref{tab: elements}.} 
\label{fig: Si_eq}
\end{center}
\end{figure}

\begin{figure}
\begin{center}
\resizebox{\hsize}{!}{\includegraphics[angle=0]{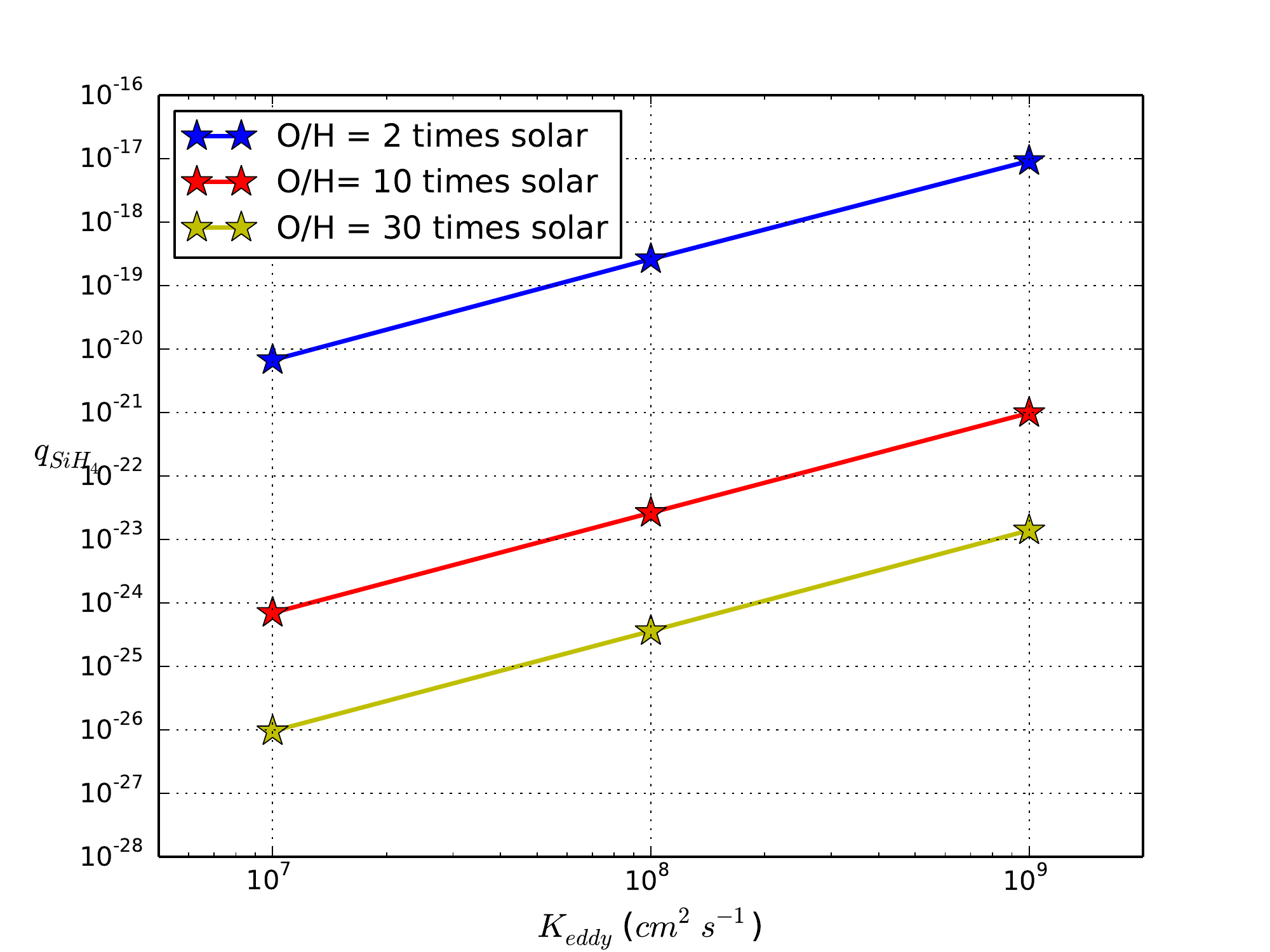}}
\caption{Predicted mixing ratio of SiH4 as a function of the vertical eddy diffusion coefficient $K_{\rm eddy}$ for Saturn. Different oxygen abundances are explored. Si/H = 10 times solar in the calculation.} 
\label{fig: SiH4}
\end{center}
\end{figure}

\begin{figure}
\begin{center}
\resizebox{\hsize}{!}{\includegraphics[angle=0]{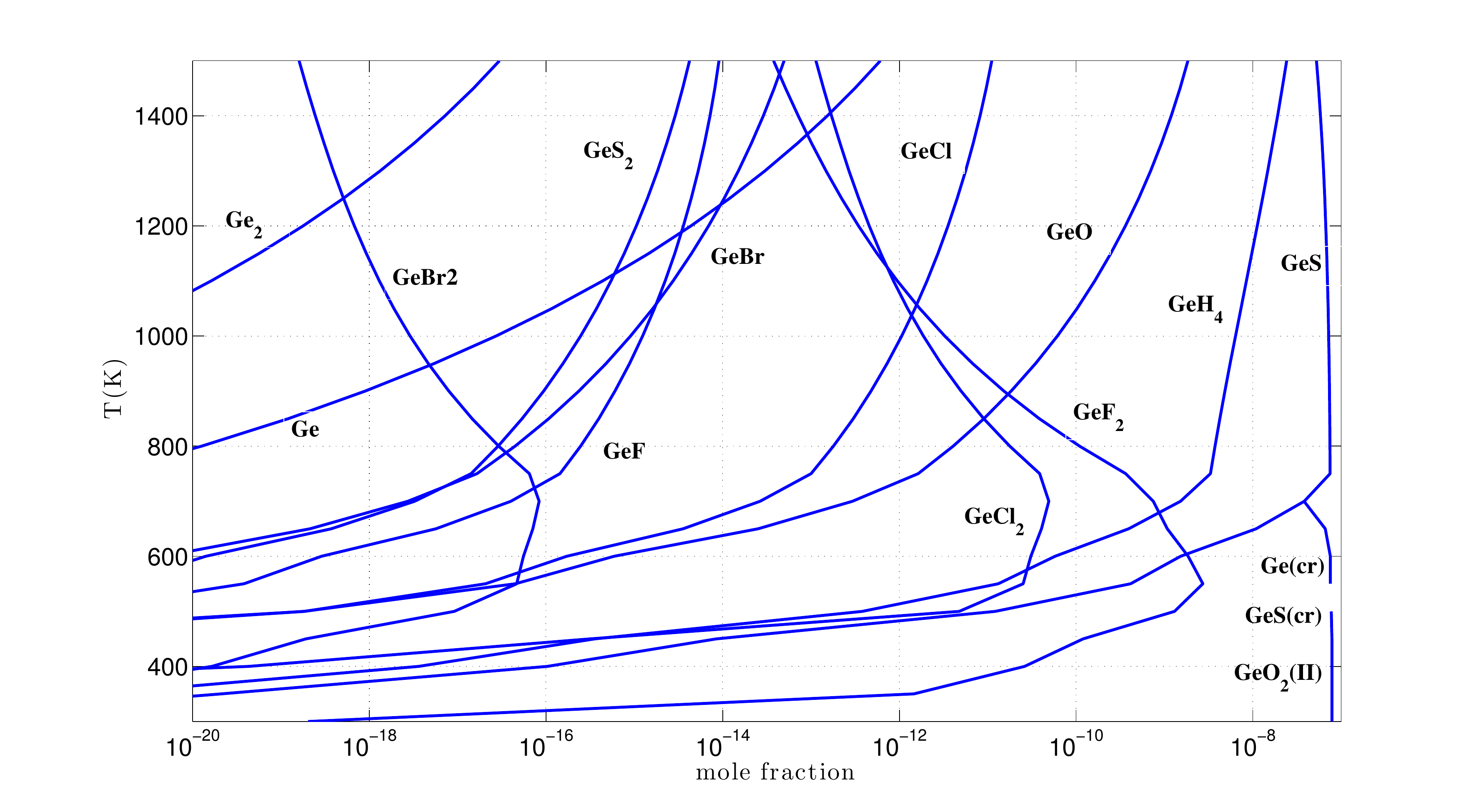}}
\caption{Equilibrium abundances of germanium containing species along Saturn's adiabat computed using NASA CEA code. We use Ge/H = 10 times solar, and O/H = 10 times solar. Other elemental abundances are also assumed to be solar except He, C, N, S, P, Si, and As, which are listed in table \ref{tab: elements}.} 
\label{fig: Ge_eq}
\end{center}
\end{figure}

\begin{figure}
\begin{center}
\resizebox{\hsize}{!}{\includegraphics[angle=0]{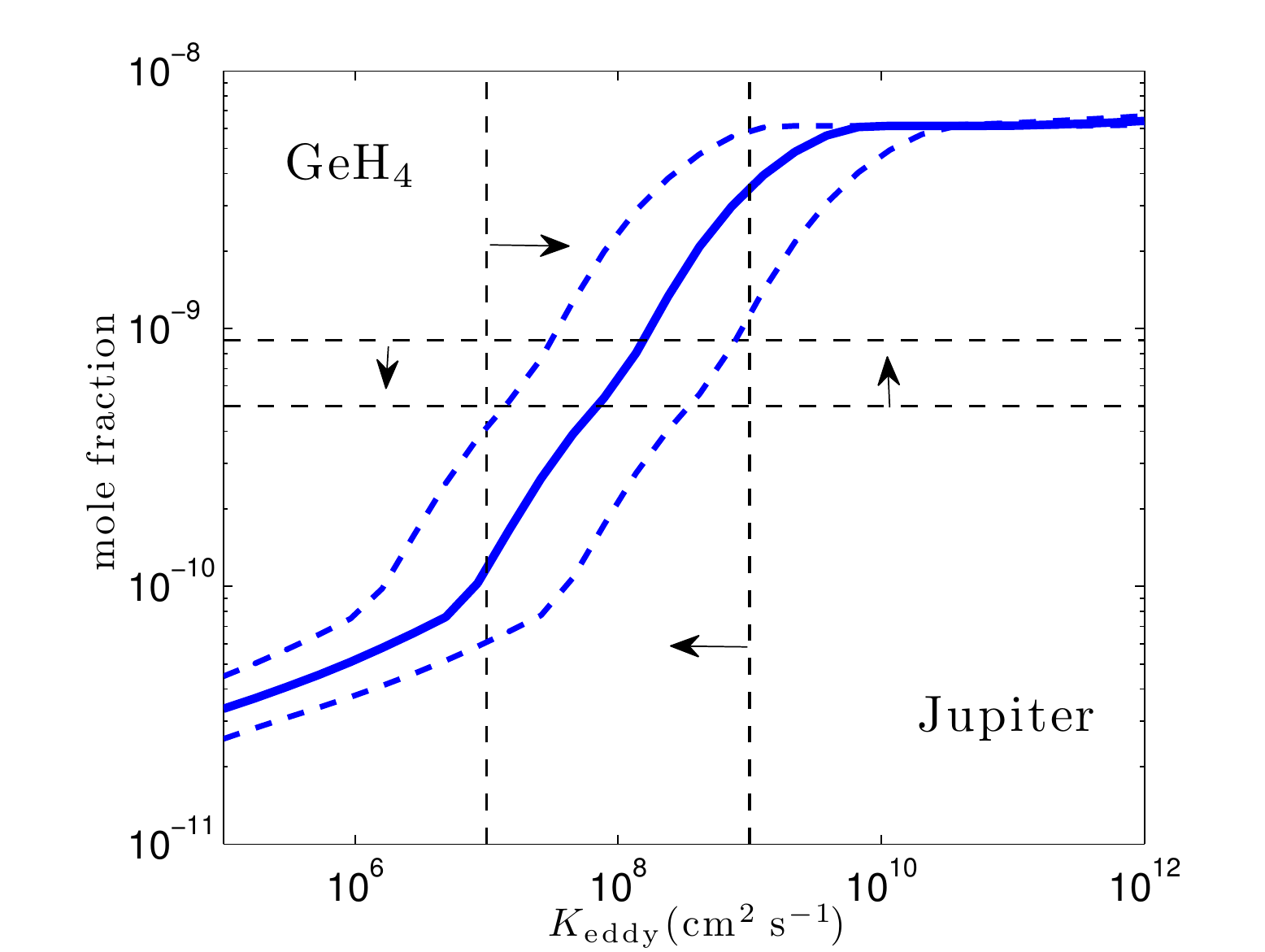}}
\caption{Predicted mole fraction of GeH$_4$ as a function of $K_{\rm eddy}$, the vertical eddy diffusion coefficient. The horizontal dashed lines show the range of observed mole fractions of GeH$_4$ \citep{Bjoraker86}. The vertical dashed lines show the plausible range of $K_{\rm eddy}$ for Jupiter. The blue dashed curves correspond to
a factor of 5 uncertainty on both sides of the rate coefficient (total factor of 25).} 
\label{fig: GeH4_J}
\end{center}
\end{figure}

\begin{figure}
\begin{center}
\resizebox{\hsize}{!}{\includegraphics[angle=0]{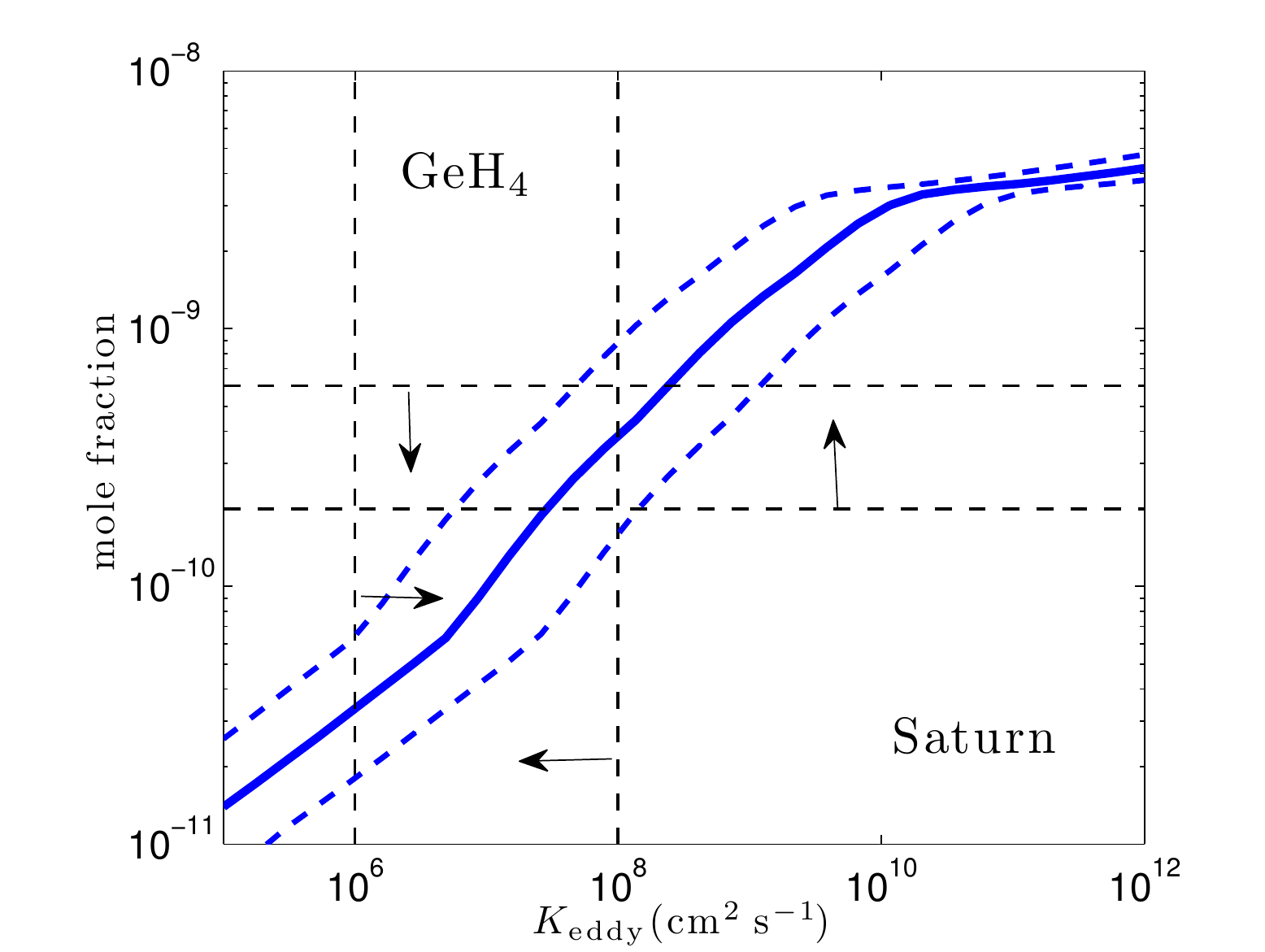}}
\caption{Predicted mole fraction of GeH$_4$ as a function of $K_{\rm eddy}$, the vertical eddy diffusion coefficient. The horizontal dashed line shows the observed range for $X_{\rm GeH_4}$, which is $4\pm2 \times 10^{-10}$ \citep{Noll88b}. The vertical dashed lines show the plausible range of $K_{\rm eddy}$ for Saturn. The blue dashed curves correspond to
a factor of 5 uncertainty on both sides of the rate coefficient (total factor of 25).} 
\label{fig: GeH4_S}
\end{center}
\end{figure}

\begin{figure}
\begin{center}
\resizebox{\hsize}{!}{\includegraphics[angle=0]{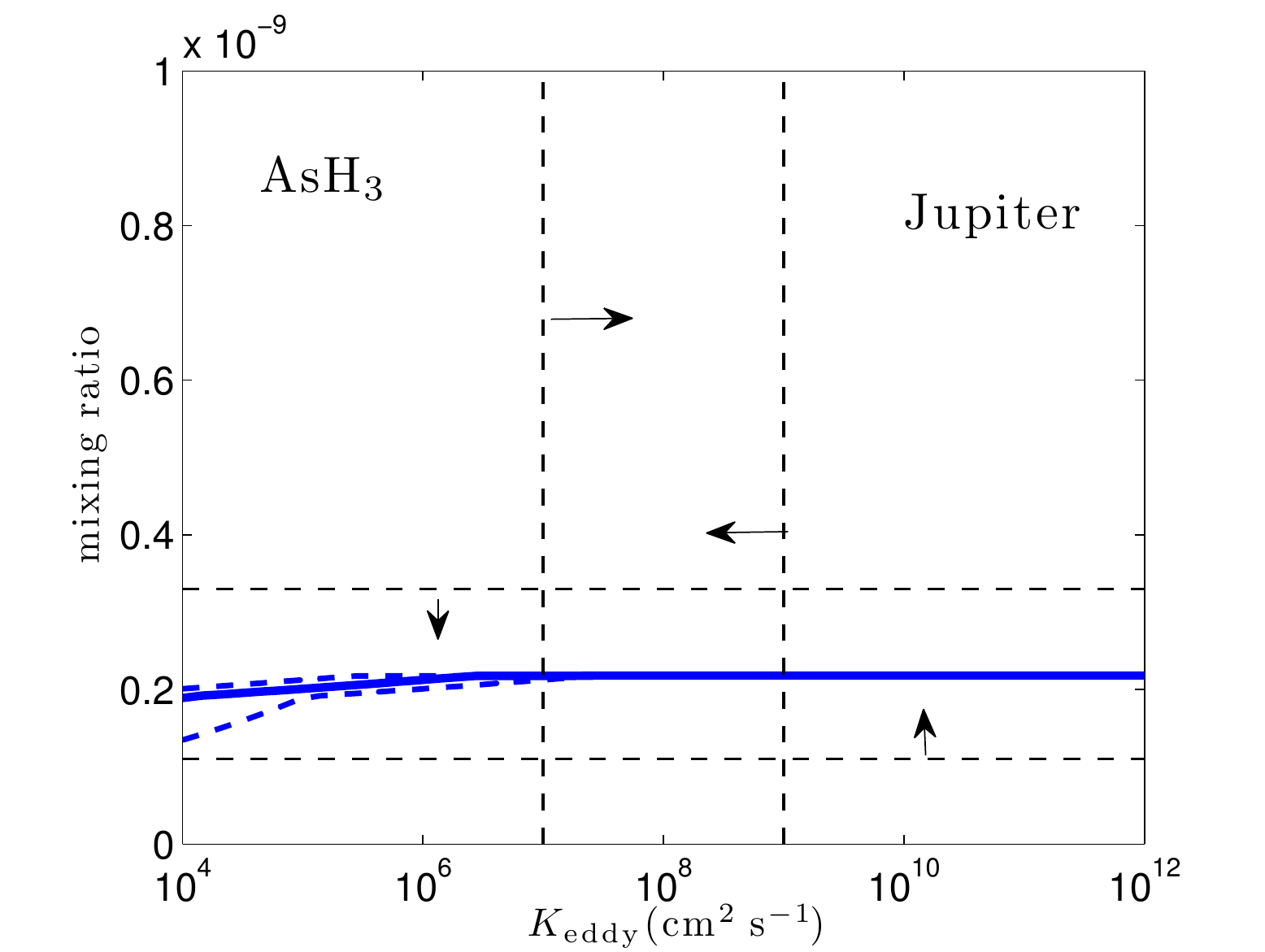}}
\caption{The predicted mixing ratio of AsH$_3$ as a function of $K_{\rm eddy}$, the vertical eddy diffusion coefficient. The horizontal dashed lines show the observed AsH$_3$ mixing ratio, $q_{\rm AsH_3}$ = 2.2$\pm$1.1$\times$10$^{-10}$ \citep{Noll90}. The vertical dashed lines show the plausible range of $K_{\rm eddy}$ \citep{Wang15}. The observed abundance of AsH$_3$ corresponds to 0.3$\sim$0.8 times solar As/H ratio.} 
\label{fig: AsH3_J}
\end{center}
\end{figure}

\begin{figure}
\begin{center}
\resizebox{\hsize}{!}{\includegraphics[angle=0]{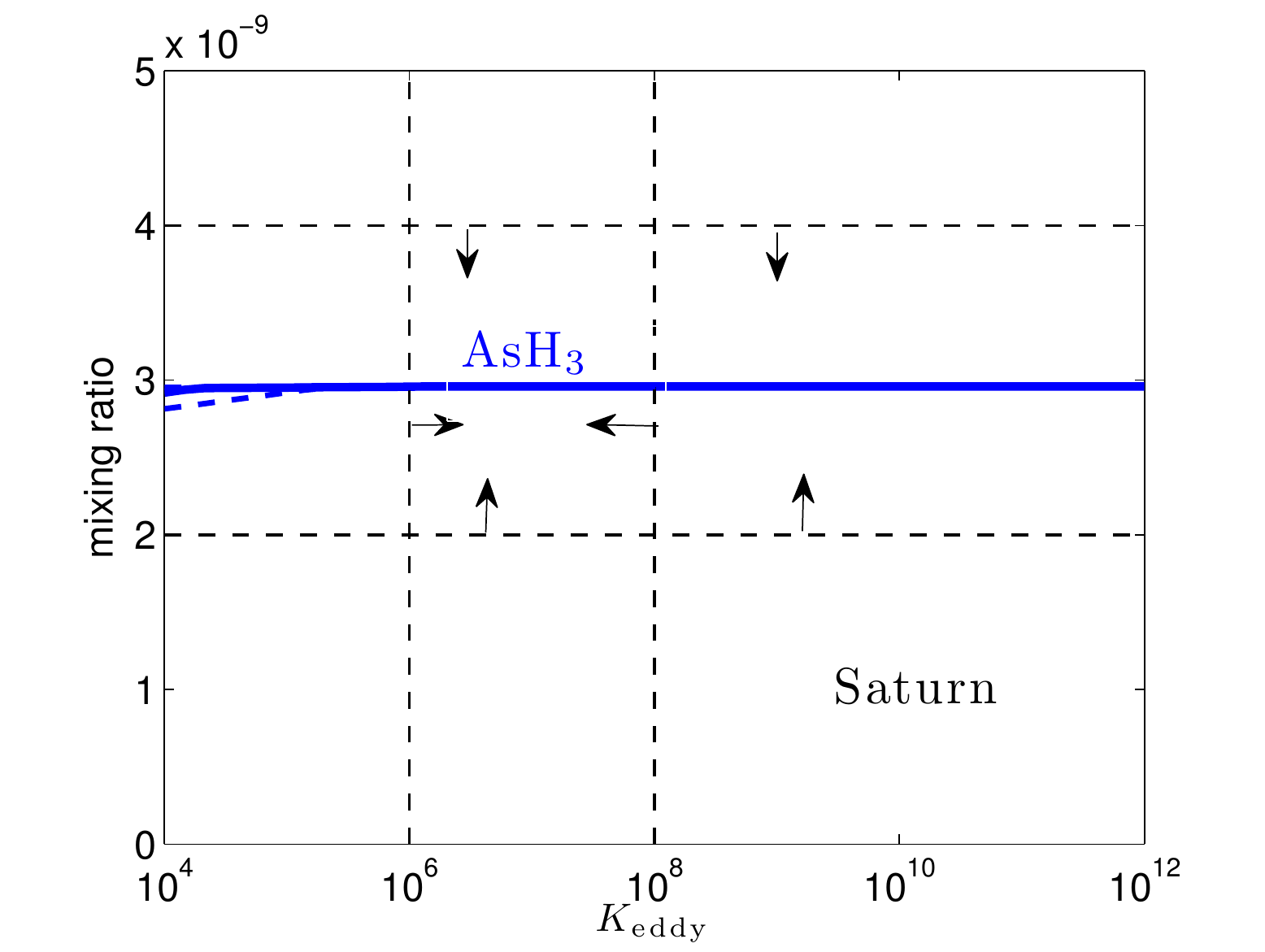}}
\caption{The predicted mixing ratio of AsH$_3$ as a function of $K_{\rm eddy}$, the vertical eddy diffusion coefficient. The horizontal dashed lines show the observed AsH$_3$ mixing ratio, $q_{\rm AsH_3}$ = 3$\pm$1$\times$10$^{-9}$ \citep{Bezard89, Noll89, NL91}. The vertical dashed lines show the plausible range of $K_{\rm eddy}$ \citep{Wang15}. The observed abundance of AsH$_3$ corresponds to 5$\sim$10 times enrichment relative to solar.} 
\label{fig: AsH3_S}
\end{center}
\end{figure}

\begin{figure}
\begin{center}
\resizebox{\hsize}{!}{\includegraphics[angle=0]{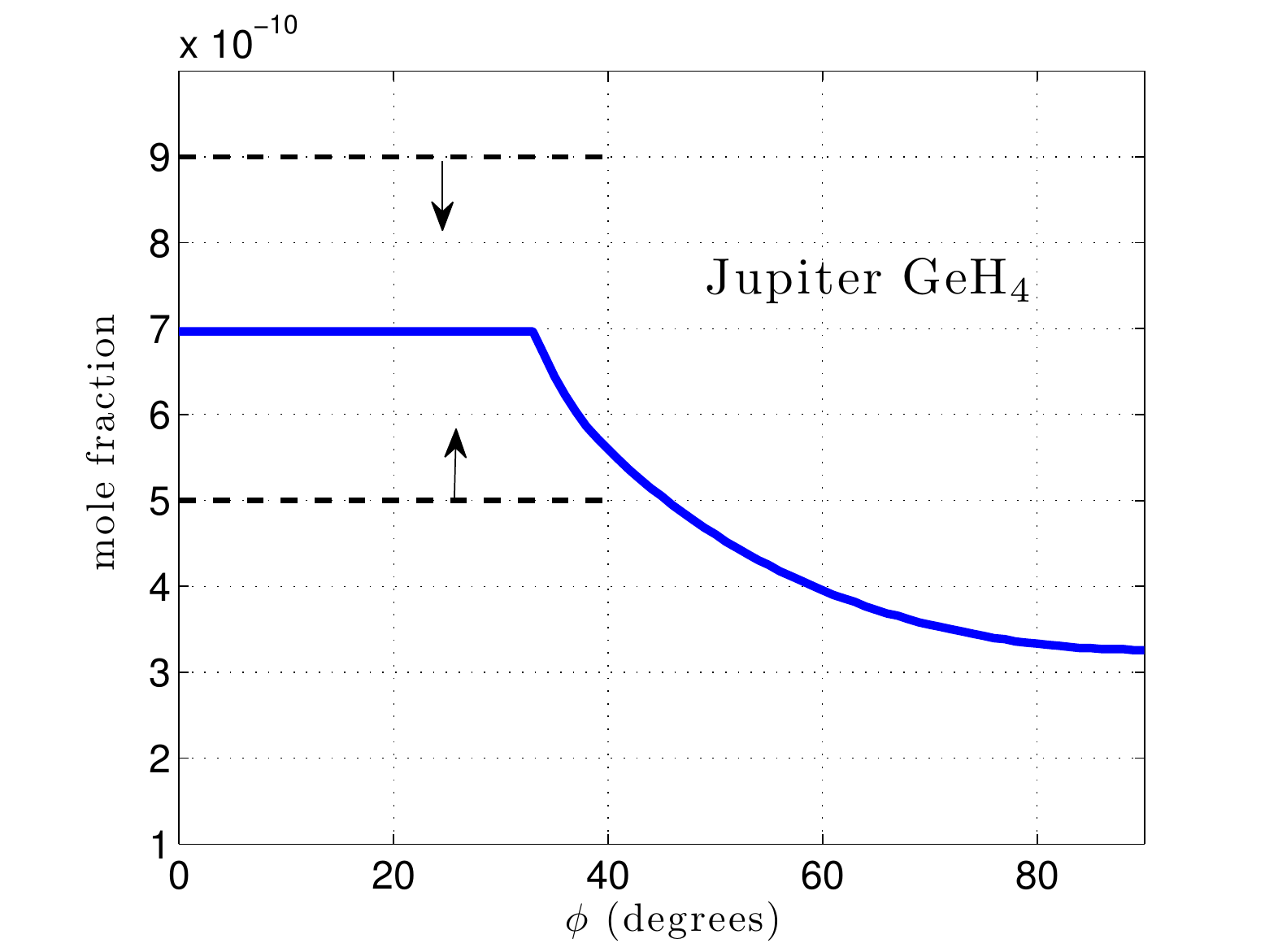}}
\caption{Mole fraction of GeH$_4$ as a function of the latitude for Jupiter computed using the germanium chemical model. The horizontal profile of the  vertical eddy diffusion coefficient used in the calculation is from \citet{Wang15}. The horizontal dashed lines show the average mole fractions of GeH$_4$ over longitude and over latitude between -40$^{\circ}$ and 40$^{\circ}$, corresponding to a value of 7$\pm$2$\times$10$^{-10}$ \citep{Bjoraker86}. } 
\label{fig: GeH4_J_horizontal}
\end{center}
\end{figure}

A Saturn entry probe with a mass spectrometer on board will be able to make in-situ measurements of the composition of Saturn, including the abundances of various disequilibrium species at a few bars level. The mass spectrometer is expected to have much higher resolution than the one on the Galileo entry probe. Therefore, it has the capability to make more precise and sensitive measurements. The probe may not be able to descend below the water cloud deck, and therefore cannot really determine the deep (global) water abundance. However, as is shown by Fig. \ref{fig: carbon_saturn}, if both C$_2$H$_6$ and CO can be measured, useful constraints can be placed on both the eddy diffusion coefficient and the deep water abundance. The mixing ratio of C$_2$H$_6$ on Saturn is predicted to be about $1\times10^{-9}$, and the mixing ratio of the tropospheric CO is predicted to be below $1\times10^{-9}$. Therefore, any instrument must be sensitive enough to measure sub-ppb level of mixing ratios. In addition, since CO and N$_2$ have nearly identical molecular weight, the payload must resolve the ppb level CO from the ppm level N$_2$. 
In \citet{Wang15}, we predicted a higher eddy diffusion coefficient near the equator and a decreasing eddy diffusion coefficient at higher latitudes. Therefore, we expect CO and C$_2$H$_6$ to be higher in abundance near the equator. The probe entry site should be preferentially near the equator in order to maximize the possibility of detecting CO and C$_2$H$_6$.
The mixing ratio of GeH$_4$ is predicted to be a few times 10$^{-10}$, and is a sensitive function of the eddy diffusion coefficient. The measurement of GeH$_4$ can add another constraint on the eddy diffusion coefficient.   

The Juno mission measurements provide a potentially important synergy with the Saturn Probe measurements. By measuring both the deep water abundance and disequilibrium species, it will be possible  to determine both the deep oxygen abundance and the magnitude of vertical eddy mixing. The latter determination is especially robust if the disequilibrium species are measured as a function of latitude \citep{Wang15}. Having a determination of the Jovian value of the eddy mixing will provide a useful constraint on that for Saturn, which will help lift the degeneracy between deep water abundance and vertical mixing from Saturn Probe measurements. Alternatively, if both CO and C$_2$H$_6$ can be measured by Saturn Probe, then an independent determination of the eddy mixing from those measurements will allow comparison of the vertical dynamics from the hundreds of bars pressure level upward on the two planets. All of this depends on the ability to resolve the uncertainties in the kinetics.  

\subsection{Uncertainties in the kinetics}
\begin{figure}
\begin{center}
\resizebox{\hsize}{!}{\includegraphics[angle=0]{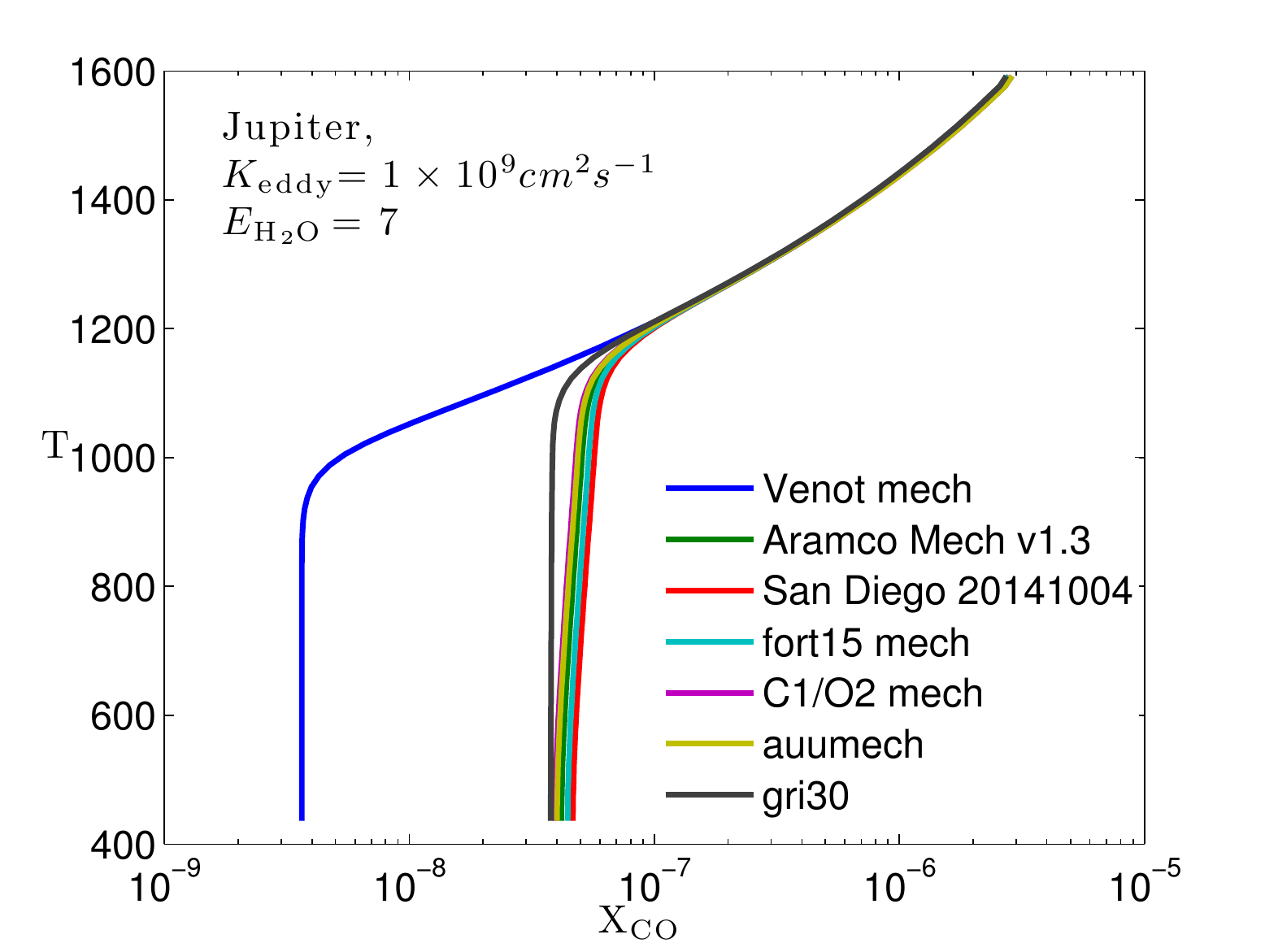}}
\caption{Comparisons on the the predicted mixing ratios of CO among various reaction networks from the combustion community. All the reaction networks shown here have been validated against many combustion experiments. } 
\label{fig: mech_comp}
\end{center}
\end{figure}

The uncertainties in the kinetics affect the constraints on both the eddy diffusion coefficient and the deep water abundance. In the planetary science community, 
various reaction networks have been developed for modeling many kinds of atmospheres, such as the atmosphere of Jupiter \citep{Visscher10}, hot Jupiters \citep[e.g.,][]{Moses11, Venot12, MK14}, terrestrial exoplanets \citep[e.g.,][]{HS14} and brown dwarfs \citep{ZM14}. These reaction networks are successful in predicting the presence of major species in the atmospheres under a wide range of temperature and pressure conditions. However, the predicted mixing ratios are subject to large errors due to the large uncertainties on individual rate coefficients in the network. As an example, the CO/CH$_4$ chemistry has been studied for decades \citep[e.g.,][]{PB77, Yung88, VM11}, yet some uncertainty still remains because of reaction coefficients under debate \citep[e.g.][]{Moses14}. The accuracy of most kinetic networks have yet to be tested by experiments except the one of \citet{Venot12}. The \citet{Venot12} network is derived from one in the combustion industry and has been validated against various combustion experiments. So is this reaction network accurate in modeling the atmospheres of giant planets? One concern is that the combustion experiments are usually conducted under carbon and oxygen rich environment, however, the atmospheres of giant planets are extremely hydrogen-rich (Moses, personal communication, 2015). We investigated several other reaction networks developed in the combustion community. The reaction networks we investigated are: the ``Aramco Mech v1.3" from \citet{Metcalfe13}; the ``San Diego Mechanism version 20141004" (\underline{http://web.eng.ucsd.edu/mae/groups/combustion/mechanism.html}); the ``C1/O$_2$ mechanism" from \citet{Li07}; the ``fort15 mech" from \citet{Sung98}; the ``aaumech" from \citet{Zabetta08}; and the ``GRI-30 mech" from \underline{http://www.me.berkeley.edu/gri\_mech/}. All the reaction networks listed above have been validated against many combustion experiments. We applied these reaction networks to Jupiter's atmosphere and compared the predicted CO mixing ratios in Fig. \ref{fig: mech_comp}. The predicted CO mixing ratios do not agree with each other, especially the \citet{Venot12} network, which is predicting a much lower value than other networks. The source of the difference is that no other networks include the channel H + CH$_3$OH $\leftrightarrow$ CH$_3$ + H$_2$O, which means this channel is not important for oxidation in an oxygen-rich environment. However, if this channel is indeed as fast as that measured in \citet{Hidaka89} as used by the \citet{Venot12} network, it can be crucial in determining the whole CO/CH$_4$ conversion rate. The comparison in Fig. \ref{fig: mech_comp} does not imply the \citet{Venot12} network is wrong, but illustrates the point that networks validated under oxygen rich environments can give different results when applied to a hydrogen-rich environment. Ideally, networks should be tested by experiments conducted under hydrogen-rich conditions in order to improve their accuracy. In this paper, due to the unresolved uncertainties in kinetics, we considered two extreme cases for CO reaction networks. The \citet{Moses11} reaction network represents the slowest pathway for CO destruction, while the \citet{Venot12} model represents the fastest pathway for CO destruction. In order to have more precise predictions on CO, C$_2$H$_6$ and GeH$_4$, a few reactions need to be further investigated experimentally or theoretically to determine their rate coefficients. For a tighter CO mixing ratio prediction, the reaction rate coefficient of H + CH$_3$OH $\leftrightarrow$ CH$_3$ + H$_2$O should be experimentally studied under conditions relevant to Jupiter and compared with the theoretical estimate by Moses et al. (2011), and the experimental estimate by \citet{Hidaka89}. For a tighter C$_2$H$_6$ mixing ratio prediction, the rate coefficient of CH$_3$ + CH$_3$ + M $\leftrightarrow$ C$_2$H$_6$ + M under high pressure should be better determined. For a better GeH$_4$ mixing ratio prediction, the rate coefficient of GeH$_2$ + H$_2$S $\leftrightarrow$ HGe=SH + H$_2$ should be determined.    
The arsenic chemistry in a hydrogen rich environment should be studied to tighten the constraints on the total As abundance. 

\section{Conclusions}   

In this paper, we used a diffusion kinetic code developed in \citet{Wang15} to predict the abundances of various disequilibrium species on Jupiter and Saturn with updated thermodynamic and kinetic data. The dependence on the vertical eddy diffusion coefficient and the deep water abundance have been explored. We summarize our simulation results below. 
\begin{itemize}
\item{We find C$_2$H$_6$ is a useful tracer for the deep eddy diffusion coefficient. The degeneracy between high (low) eddy diffusion coefficient and high (low) deep water abundance from CO constraints can be broken by adding the constraint by C$_2$H$_6$. }
\item{We find PH$_3$ is converted to H$_3$PO$_4$ instead of P$_4$O$_6$ as in previous studies. We identified a new chemical pathway based on a H/P/O reaction network. The PH$_3$ abundance is predicted to be insensitive to either the eddy diffusion coefficient or the deep water abundance unless $E_{\rm H_2O}$ is higher than 20.  }
\item{We confirm that SiH$_4$ is not expected in the troposphere of either Jupiter or Saturn based on a H/Si/O reaction network. A new chemical pathway for SiH$_4$/MgSiO$_3$ (s) conversion is proposed. }
\item{We propose a new chemical pathway for GeH$_4$ destruction. The GeH$_4$ abundance is predicted to be a sensitive function of the eddy diffusion coefficient. }
\item{We confirm that the element As is primarily sequestered in AsH$_3$ in Jupiter and Saturn's atmosphere by exploring a new chemical pathway for AsH$_3$ destruction. }
\item{Since the eddy diffusion coefficient is predicted by theoretical models to be latitudinally dependent, we predict the tropospheric abundances of CO, C$_2$H$_6$, CO$_2$, and GeH$_4$ to have latitudinal variations, and the tropospheric abundances of PH$_3$ and AsH$_3$ to have no latitudinal variations. }
\end{itemize}
Juno can provide multiple constraints on the eddy diffusion coefficient from its measurement of disequilibrium species by JIRAM and its measurement of the deep water abundance from the microwave radiometer. 
A probe with a mass spectrometer sensitive enough to detect sub-ppb level of CO and C$_2$H$_6$ can place constraints on both the deep diffusion coefficient and the deep water abundance on Saturn. A probe should be sent to equatorial latitude to maximize the probability of detecting disequilibrium species. 
The predictions on the disequilibrium chemistry are limited by the uncertainty in kinetics. Several reactions are worth further investigations to reduce the uncertainties and they are H + CH$_3$OH $\leftrightarrow$ CH$_3$ + H$_2$O, CH$_3$ + CH$_3$ + M $\leftrightarrow$ C$_2$H$_6$ + M, and GeH$_2$ + H$_2$S $\leftrightarrow$ HGe=SH + H$_2$.

\section*{Acknowledgments}
We thank O. Venot for providing her reaction network. We are most grateful to B. B\'{e}zard and T. Cavali\'{e} for their careful reviews of the paper. Support from Juno project is gratefully acknowledged. This work has been partly carried out thanks to the support of the A*MIDEX project (n\textsuperscript{o} ANR-11-IDEX-0001-02) funded by the ``Investissements d'Avenir'' French Government program, managed by the French National Research Agency (ANR). D. W. and J. L. are supported by the Juno project. 



\bibliographystyle{elsarticle-harv}
\bibliography{master}


\clearpage
\begin{table}[h]
\centering \caption{Observed mixing ratios of some disequilibrium species}
\begin{center}
\begin{tabular}{lcccc}
\hline \hline
 & \multicolumn{2}{c}{Jupiter} & \multicolumn{2}{c}{Saturn}  \\
\hline
& $q$  & references & $q$  & references\\
\hline
CO & 1.0$\pm$0.2$\times$10$^{-9}$  & \citet{Bezard02} & $<$1.0$\times$10$^{-9}$  &  \citet{Cavalie09}    \\
PH$_3$ & 8$\pm$1$\times$10$^{-7}$ &  \citet{Irwin98} & 4$\pm$1$\times$10$^{-6}$ & \citet{Fletcher11}    \\
SiH$_4$ & $<$2.5$\times$10$^{-9}$  & \citet{Treffers78} & $<$2$\times$10$^{-10}$   & \citet{NL91}   \\
GeH$_4$ & 7$\pm$2$\times$10$^{-10}$&  \citet{Bjoraker86} & 4$\pm$2$\times$10$^{-10}$ & \citet{Noll88b}        \\
AsH$_3$ & 2.2$\pm$1.1$\times$10$^{-10}$  & \citet{Noll90} & 3$\pm$1$\times$10$^{-9}$   & \citet{NL91}       \\
\hline
\end{tabular}
\end{center}
\label{tab: compositions}
\end{table}

\begin{table}[h]
\centering \caption{Elemental abundances used in the simulation for Jupiter and Saturn.}
\begin{center}
\begin{tabular}{lcccccc}
\hline \hline
 & \multicolumn{3}{c}{Jupiter} & \multicolumn{3}{c}{Saturn}  \\
\hline
& $q$  & E  & references & $q$  & E & references\\
\hline
He & 0.157 & 0.920 & \citet{Niemann98} & 0.135 & 0.794 & \citet{CG00}       \\
C & 2.37$\times$10$^{-3}$ & 4.4 & \citet{Wong04} & 5.33$\times$10$^{-3}$ & 9.91  & \citet{Fletcher09b}        \\
N & 6.64$\times$10$^{-4}$ & 4.92 & \citet{Wong04} & 4.54$\times$10$^{-4}$ & 3.36 & \citet{Fletcher11}       \\
O & \multicolumn{3}{c}{variable in the simulation} &  \multicolumn{3}{c}{variable in the simulation}  \\
S & 8.90$\times$10$^{-5}$ & 3.38 & \citet{Wong04} & 3.76$\times$10$^{-4}$ & 14.3  &   \citet{BS89}      \\
P & 7$\times$10$^{-7}$ & 1.4 & \citet{Irwin98} & 4$\times$10$^{-6}$ & 7.8 & \citet{Fletcher11}    \\
Si & 2.85$\times$10$^{-4}$ & 4.4 & assumed & 6.47$\times$10$^{-4}$ & 10  &  assumed       \\
Ge & 3.93$\times$10$^{-8}$& 4.4 & assumed & 8.93$\times$10$^{-8}$ & 10 & assumed        \\
As & 2.2$\times$10$^{-10}$ & 0.55 & \citet{Noll90} & 3.0$\times$10$^{-9}$ & 7.5  & \citet{NL91}      \\
\hline
\end{tabular}
\end{center}
\label{tab: elements}
\end{table}

\clearpage

\clearpage

\clearpage

\clearpage

\clearpage

\clearpage

\clearpage

\clearpage

\clearpage

\clearpage
   
\clearpage   

\clearpage

\clearpage

\clearpage

\clearpage

\clearpage

\clearpage

\clearpage
\appendix
  
\section{Chemical pathway for CO/CH$_4$ conversion}

In this appendix, we introduce our method of identifying the main chemical pathway for CO/CH$_4$ conversion.    
The main chemical pathway for CO/CH$_4$ conversion is automatically identified from the C/N/O/H reaction network by examining the conversion rate between different carbon bearing species using \textit{Cantera}. As an example, consider the conversion between CH$_3$O and CH$_3$OH, a non-exhaustive list of relevant reactions are

\begin{subequations}\label{eqn: CH3O-CH3OH}
\begin{align}
& \textrm{CH}_3\textrm{O} +  \textrm{HCO} \leftrightarrow \textrm{CH}_{3}\textrm{OH} + \textrm{CO}, \\
& \textrm{CH}_3\textrm{O} + \textrm{H}_{2}\textrm{CO} \leftrightarrow \textrm{CH}_3\textrm{OH} + \textrm{HCO}, \\
& \textrm{CH}_3\textrm{O} + \textrm{CH}_3\textrm{O} \leftrightarrow \textrm{CH}_3\textrm{OH} + \textrm{H}_{2}\textrm{CO}, \\
& \textrm{CH}_3\textrm{O} + \textrm{CH}_2\textrm{OH} \leftrightarrow \textrm{CH}_3\textrm{OH} + \textrm{H}_{2}\textrm{CO}, \\  
& \textrm{CH}_3\textrm{O} + \textrm{H}_2 \leftrightarrow \textrm{CH}_3\textrm{OH} + \textrm{H},   \\
& .......
\end{align}
\end{subequations}

\begin{figure}
  \begin{subfigure}[b]{.5\textwidth}
       \includegraphics[width=\textwidth]{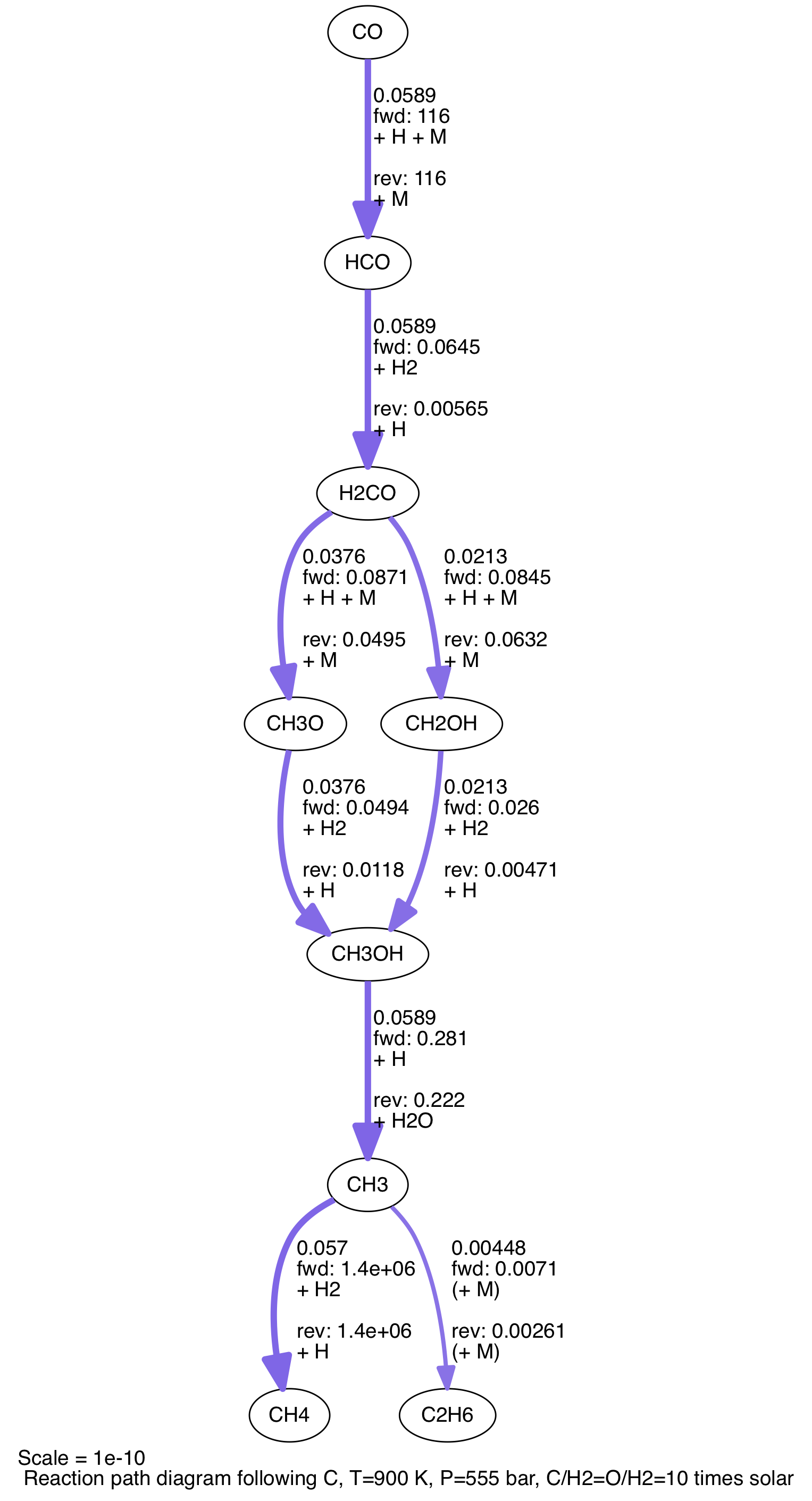}
  \end{subfigure}
  \begin{subfigure}[b]{.5\textwidth}
      \includegraphics[width=\textwidth]{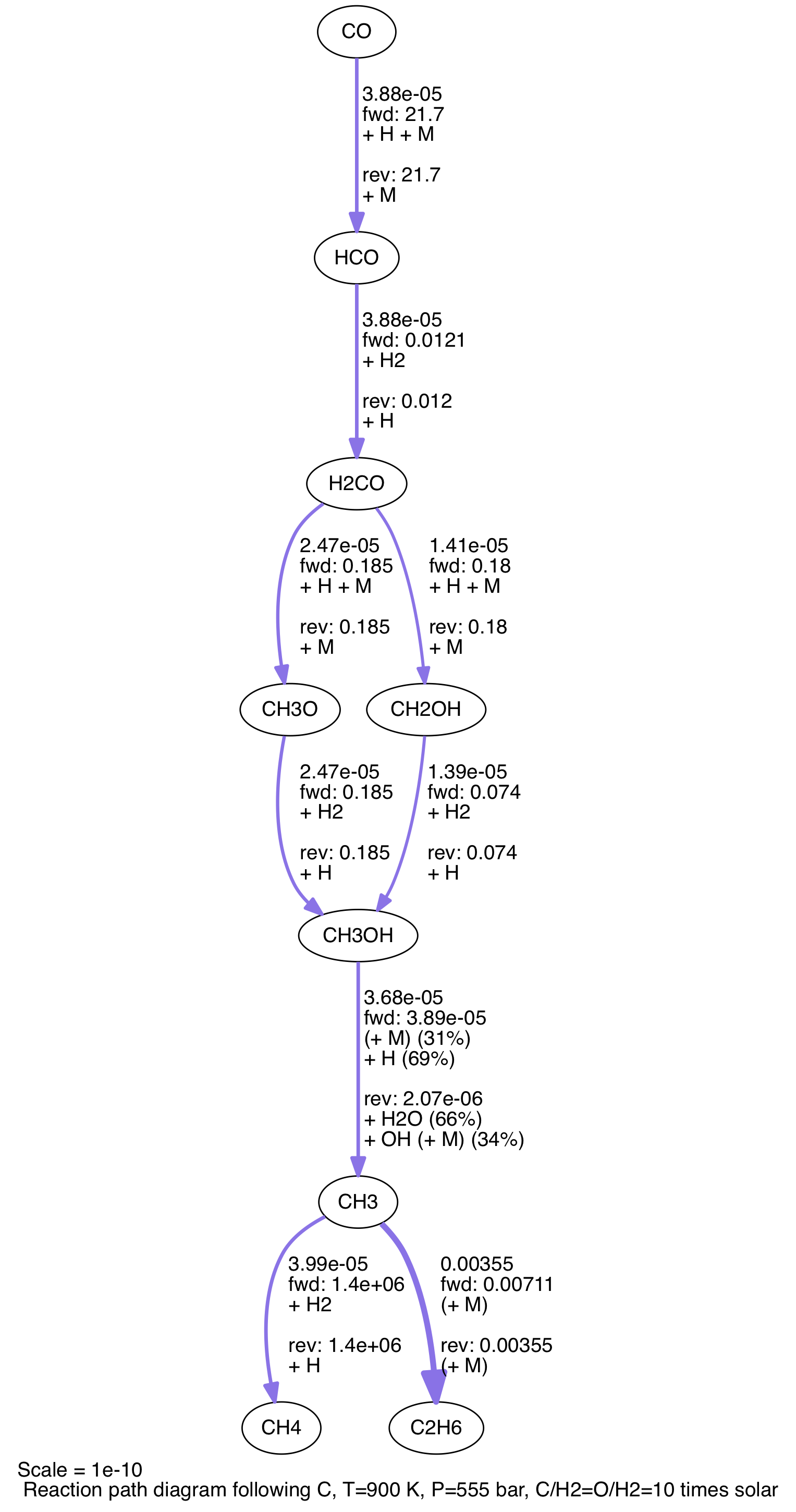}
  \end{subfigure}
\caption{Reaction path diagram for carbon bearing species generated by \textit{Cantera}. The left figure is generated using the C/N/O/H reaction network A, 
and the right figure is generated using the network B. 
The composition and temperature-pressure condition resemble the CO quench level of Saturn ($\sim$ 900 K, 550 bars). The arrows show the flow directions of element carbon, and the labels show the net flux (in the unit of mole cm$^{-3}$ s$^{-1}$) of element carbon as well as the forward flux and backward flux. The reactions responsible for the forward and backward flux are also labeled. All the carbon bearing species are considered, but only fluxes above a threshold are shown in the diagram.} 
\label{fig: path_C}
\end{figure}

The total conversion rate from CH$_3$O to CH$_3$OH, the inverse conversion rate, and the net conversion rate are computed by summing all the rates of each individual reactions. All pairs of carbon bearing species are considered in the computation by \textit{Cantera}. Then the conversion rates are displayed with a diagram generated using a method called \textit{ReactionPathDiagram} in \textit{Cantera}. The conversions with rates higher than a threshold are summarized in the diagram. Specifically, we start our simulation from a mixture of Saturn composition gas under the condition of Saturn's CO quench level ($\sim$ 900 K, 500 bars), with CO abundance slightly higher than the equilibrium abundance. The system automatically evolves to the chemical equilibrium state, and some fraction of CO is converted to CH$_4$. This conversion process can be viewed as  
a flow of carbon element from CO to CH$_4$ (and other species) across the reaction network. The flow channel with the fastest rate is the main chemical pathway connecting CO and CH$_4$. 
In Fig. \ref{fig: path_C}, we show two reaction path diagrams, one using the network A and the other using the network B. The net flux of element carbon
from one species to another species are labeled in the diagrams. By following the path with the highest net flux from CO to CH$_4$, we find the main chemical pathway for CO/CH$_4$ conversion as 

\begin{subequations}\label{eqn: C_pathway_A}
\begin{align}
& \textrm{CO} + \textrm{H} + \textrm{M} \leftrightarrow \textrm{HCO} + \textrm{M},    \\
& \textrm{HCO} +  \textrm{H}_{2} \leftrightarrow \textrm{H}_{2}\textrm{CO} + \textrm{H}, \\
& \textrm{H}_{2}\textrm{CO} + \textrm{H} + \textrm{M} \leftrightarrow \textrm{CH}_3\textrm{O} + \textrm{M}, \\
& \textrm{CH}_3\textrm{O} + \textrm{H}_2 \leftrightarrow \textrm{CH}_3\textrm{OH} + \textrm{H}, \\
& \textrm{H}_2\textrm{CO} + \textrm{H} + \textrm{M} \leftrightarrow \textrm{CH}_2\textrm{OH} + \textrm{M}, \\  
& \textrm{CH}_2\textrm{OH} + \textrm{H}_2 \leftrightarrow \textrm{CH}_3\textrm{OH} + \textrm{H},   \\
& \textrm{CH}_3\textrm{OH} + \textrm{H}  \leftrightarrow \textrm{CH}_{3} + \textrm{H}_{2}\textrm{O}, \\
& \textrm{CH}_3\textrm{OH} + \textrm{M}  \leftrightarrow \textrm{CH}_{3} + \textrm{OH} + \textrm{M}, \\
& \textrm{CH}_{3} + \textrm{H}_{2}  \leftrightarrow \textrm{CH}_{4} + \textrm{H}, \\
\cline{1-2}
& \textrm{CO} + 3\textrm{H}_{2}  \leftrightarrow \textrm{CH}_{4} + \textrm{H}_{2}\textrm{O}. \tag{\ref{eqn: C_pathway_A}, net} 
\end{align}
\end{subequations}

The conversion from H$_2$CO to CH$_3$OH follows two branches.  
The two networks agree with each other on the main chemical pathway.

Among the chemical pathway, some steps are slower than other steps, which serves as the bottleneck of the pathway. The slowest steps among the main
pathway are usually called the rate determining steps. Fast steps can reach chemical equilibrium very quickly, but the slow steps are far from equilibrium. The degree of non-equilibrium can be defined in the following way. For reactions from species A to species B, we define the forward rate as $r_f$ and the 
backward rate as $r_b$. The non-equilibrium parameter can be defined as 

\begin{equation}\label{eqn: f_noneq}
f_{\rm noneq} = (r_f - r_b)/r_f, 
\end{equation}

$f_{\rm noneq}$ is close to zero for nearly equilibrium and between zero and one for non equilibrium. In table \ref{tab: C}, we show the computed value of $f_{\rm noneq}$ for all the steps on the
pathway. For the simulations using the network A, there are several steps that are far from equilibrium. Only the first step and the last step 
are close to equilibrium. This indicates that the steps HCO $\rightarrow$ H$_2$CO and H$_2$CO $\rightarrow$ CH$_3$OH have similar rate and both bottleneck the overall chemical pathway. Using a quench-level model, we find using a single rate determining step will introduce an error about 30$\sim$50$\%$ in the prediction.  For the simulations using the 
network B, the step from CH$_3$OH to CH$_3$ is the only step far from equilibrium, therefore, it is the rate determining step. Since the network B is similar to the network in Moses et al. (2011), our findings of the rate determining step is therefore the same as the one identified by \citet{Moses11} and \citet{VM11}. The different choice of rate determining step is easy to understand. Since the network A adopted a larger reaction constant for the reaction H + CH$_3$OH $\leftrightarrow$ CH$_3$ + H$_2$O, the conversion from CH$_3$OH to CH$_3$ is no not a bottleneck, and the overall rate increases.    

\clearpage

\clearpage
\begin{table}[h]
\centering \caption{Non-equilibrium parameter $f_{\rm noneq}$ for reactions on the CO/CH$_4$ chemical pathway.}
\begin{center}
\begin{tabular}{lcc}
\hline \hline
reactions & $f_{\rm noneq}$, for network A, t = 4$\times$10$^4$ s  &  $f_{\rm noneq}$, for network B, t = 4$\times$10$^6$ \\
\hline
CO $\rightarrow$ HCO &  5.1$\times$10$^{-4}$   &  1.8$\times$10$^{-6}$    \\
HCO $\rightarrow$ H$_2$CO  & 9.1$\times$10$^{-1}$  &    3.2$\times$10$^{-3}$  \\
H$_2$CO $\rightarrow$ CH$_3$O  & 4.3$\times$10$^{-1}$  &  1.3$\times$10$^{-4}$   \\
H$_2$CO $\rightarrow$ CH$_2$OH &  2.5$\times$10$^{-1}$    &   7.8$\times$10$^{-5}$  \\
CH$_3$O $\rightarrow$ CH$_3$OH  &  7.6$\times$10$^{-1}$ &    1.3$\times$10$^{-4}$   \\
CH$_2$OH $\rightarrow$ CH$_3$OH & 8.2$\times$10$^{-1}$  &    1.9$\times$10$^{-4}$  \\
CH$_3$OH $\rightarrow$ CH$_3$ &  2.1$\times$10$^{-1}$ &  9.5$\times$10$^{-1}$\\
CH$_3$ $\rightarrow$ CH$_4$ & 4.1$\times$10$^{-8}$ &  2.5$\times$10$^{-11}$\\
\hline
\end{tabular}
\end{center}
\label{tab: C}
\end{table}

\section{Chemical pathway for PH$_3$/H$_3$PO$_4$ conversion}

\begin{figure}
\begin{center}
\includegraphics[width=0.5\textwidth]{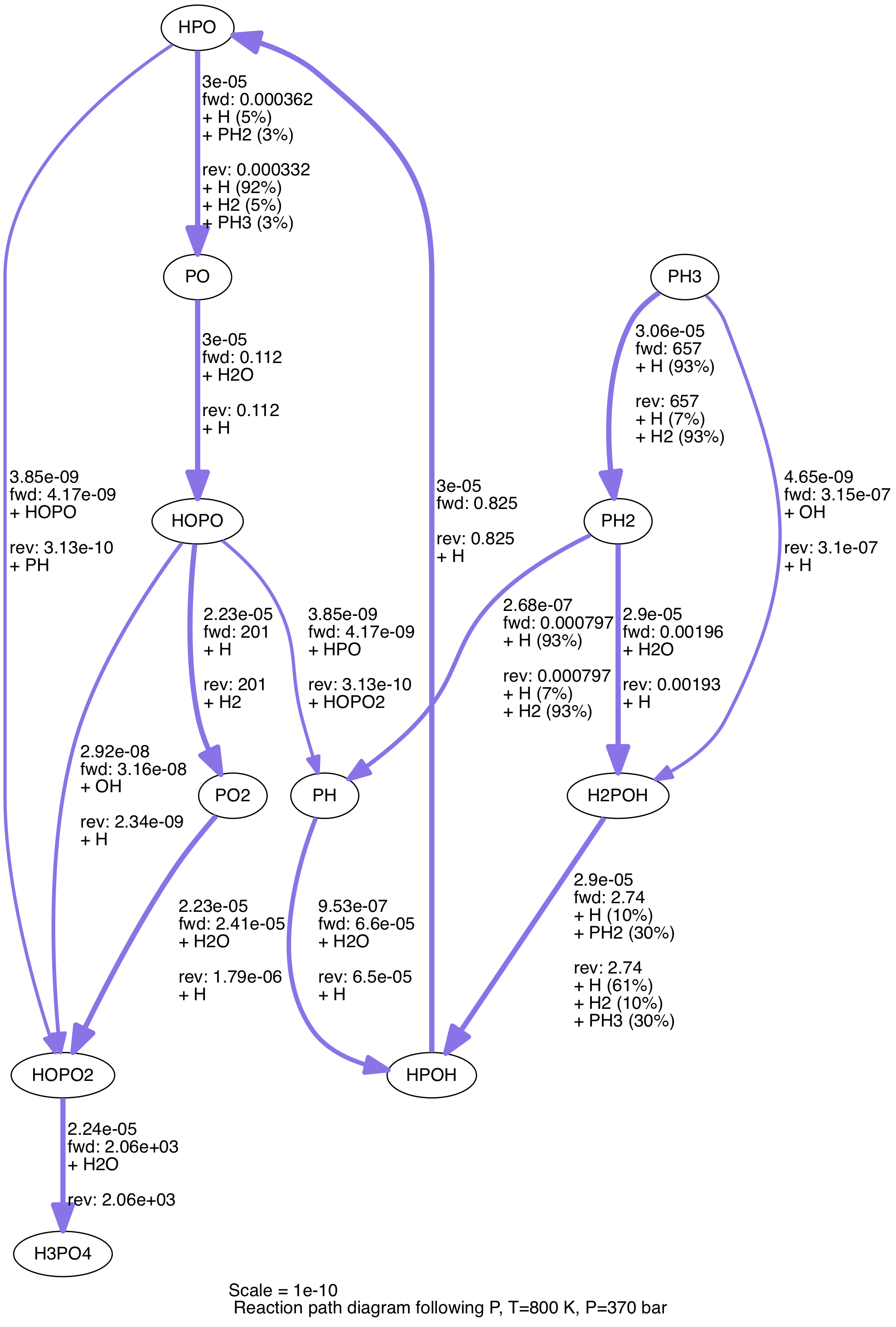}
\caption{Reaction path diagram for phosphorus bearing species generated by \textit{Cantera}. The input reaction network here is the H/P/O network described in section 3.1. 
The arrows in the diagram show the flow directions of element phosphorus. The labels show the 
net element flux (in the unit of mole cm$^{-3}$ s$^{-1}$) as well as the forward flux and the backward flux. The reactions responsible for
the forward and backward flux are also labeled. All species in the network are considered, but only species that have net flux above a threshold are shown in this figure.} 
\label{fig: path_P}
\end{center}
\end{figure}

The chemical pathway for PH$_3$ destruction is automatically identified by examining the conversion rate between all species in the reaction network. The method is detailed in the Appendix A using the CO/CH$_4$ conversion as an example. We start our simulation from a mixture with the Saturn-like composition: He/H = 0.27, O/H = 20 times solar, and P/H =  8 times solar. The temperature and pressure are held constant as the mixture chemically evolves to an equilibrium state. In Fig. \ref{fig: path_P}, we show the reaction path diagram for phosphorus bearing species generated using \textit{Cantera}. PH$_3$ is converted to H$_3$PO$_4$ bypassing a list of species: PH$_2$, H$_2$POH, HPOH, HPO, 
PO, HOPO, PO$_2$, and HOPO$_2$. In table \ref{tab: P}, we present the computed values
of $f_{\rm noneq}$ at one instant during the chemical evolution. The only step that is far from equilibrium is the step PO$_2$ $\rightarrow$ HOPO$_2$. Therefore, the rate determining reaction for PH$_3$/H$_3$PO$_4$ conversion near the PH$_3$ quench level of Saturn is PO$_2$ + H$_2$O $\leftrightarrow$ HOPO$_2$ + H. The rate determining step could be different if the temperature and pressure conditions were very different from 800 K and 370 bars. 

\clearpage
 
\clearpage
\begin{table}[h]
\centering \caption{Non-equilibrium parameter $f_{\rm noneq}$ for reactions on the PH$_3$/H$_3$PO$_4$ chemical pathway.}
\begin{center}
\begin{tabular}{lcc}
\hline \hline
reactions & $f_{\rm noneq}$, t = 1$\times$10$^8$ s   \\
\hline
PH$_3$ $\rightarrow$ PH$_2$ &  4.7$\times$10$^{-8}$      \\
PH$_2$ $\rightarrow$ H$_2$POH  & 1.5$\times$10$^{-2}$  \\
H$_2$POH $\rightarrow$ HPOH  & 1.1$\times$10$^{-5}$     \\
HPOH $\rightarrow$ HPO &  3.6$\times$10$^{-5}$     \\
HPO $\rightarrow$ PO  &  8.3$\times$10$^{-2}$   \\
PO $\rightarrow$ HOPO & 2.7$\times$10$^{-4}$    \\
HOPO $\rightarrow$ PO$_2$ &  1.1$\times$10$^{-7}$  \\
PO$_2$ $\rightarrow$ HOPO$_2$ & 9.3$\times$10$^{-1}$  \\
HOPO$_2$ $\rightarrow$ H$_3$PO$_4$ & 1.1$\times$10$^{-8}$  \\
\hline
\end{tabular}
\end{center}
\label{tab: P}
\end{table}
 
\section{Chemical pathway for SiH$_4$ destruction}

\begin{figure}
\begin{center}
\includegraphics[width=0.3\textwidth]{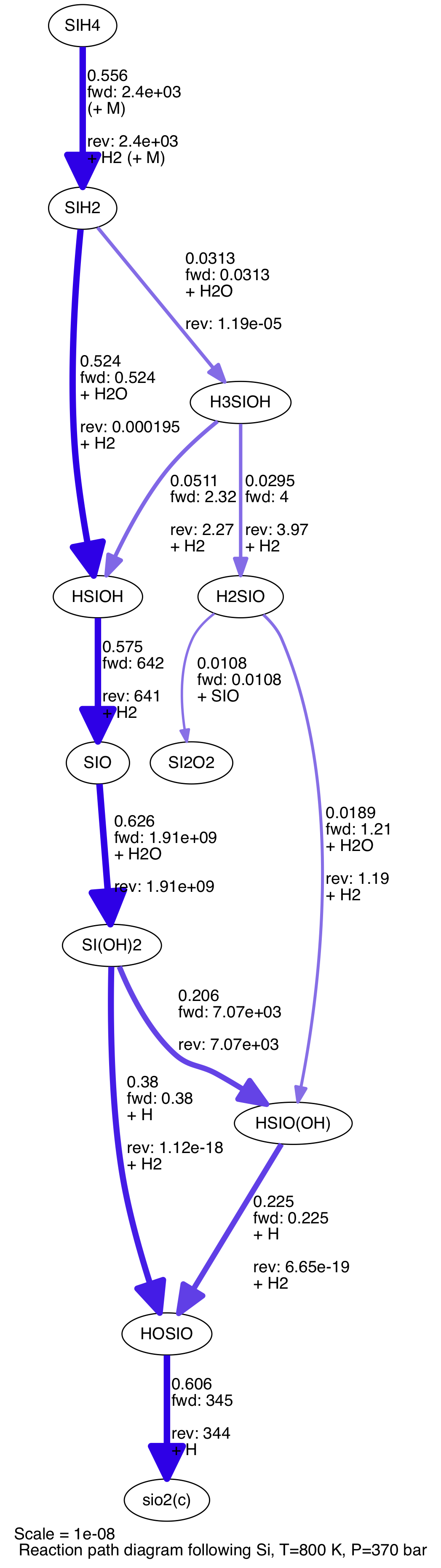}
\caption{Chemical pathway for silicon bearing species generated by \textit{Cantera}. The input reaction network here is the H/Si/O network described in section 3.1. 
The arrows show the flow directions of element silicon. The labels show the 
net flux (mole cm$^{-3}$ s$^{-1}$) of element silicon as well the forward flux and the backward flux. Also labeled are the reactions responsible for the forward and backward flux. All species in the network are considered, but only species that have net flux above a threshold are shown in the diagram.} 
\label{fig: path_Si}
\end{center}
\end{figure}

Silane (SiH$_4$) is oxidized into various species in the atmosphere of Jupiter and Saturn. The major products are MgSiO$_3$ condensates. The main chemical pathway is identified following the approach detailed in the Appendix A using CO/CH$_4$ conversion as an example. We start our simulation from a mixture with a Saturn like composition: He/H = 0.27, O/H = 10 times solar, and Si/H = 10 times solar. The temperature and pressure are held constant as the mixture evolves into
chemical equilibrium state. The reaction path diagram is shown in Fig. \ref{fig: path_Si}. The main destruction pathway of SiH$_4$ is identified from the diagram: SiH$_4$ $\rightarrow$ SiH$_2$
$\rightarrow$ HSiOH $\rightarrow$ SiO $\rightarrow$ Si(OH)$_2$ $\rightarrow$ (HOSiOOH) $\rightarrow$ HOSiO $\rightarrow$ SiO$_2$. SiO$_2$ will react with Mg(OH)$_2$ to form MgSiO$_3$ under the conditions of 
Jupiter and Saturn. In table \ref{tab: Si}, we present the computed values of $f_{\rm noneq}$ at an instant
of the chemical evolution. There are three steps that are far from equilibrium. The first step is the
conversion from SiH$_2$ to HSiOH, and the other two are both at the step from Si(OH)$_2$ to HOSiO. There is no unique rate-determining step for SiH$_4$ destruction. 

\clearpage
 
\clearpage 
\begin{table}[h]
\centering \caption{Non-equilibrium parameter $f_{\rm noneq}$ for reactions on the SiH$_4$ destruction chemical pathway.}
\begin{center}
\begin{tabular}{lcc}
\hline \hline
reactions & $f_{\rm noneq}$, t = 2$\times$10$^5$ s   \\
\hline
SiH$_4$ $\rightarrow$ SiH$_2$ &  2.3$\times$10$^{-4}$      \\
SiH$_2$ $\rightarrow$ H$_2$SiOH  & 1.0  \\
HSiOH $\rightarrow$ SiO  & 9.0$\times$10$^{-4}$     \\
SiO $\rightarrow$ Si(OH)$_2$  & 3.3$\times$10$^{-10}$     \\
Si(OH)$_2$ $\rightarrow$ HOSiO &  1.0     \\
Si(OH)$_2$ $\rightarrow$ HSiO(OH)  &  2.9$\times$10$^{-5}$   \\
HSiO(OH) $\rightarrow$ HOSiO & 1.0    \\
HOSiO $\rightarrow$ SiO$_2$(c) &  1.8$\times$10$^{-3}$  \\
\hline
\end{tabular}
\end{center}
\label{tab: Si}
\end{table}

\end{document}